\documentclass[]{aastex63}

\received{10 February 2020}
\revised{}
\accepted{}
\submitjournal{PSJ}

\shorttitle{Titan: Earth-like on the Outside, Ocean World on the Inside}
\shortauthors{MacKenzie et al.}

\begin{document}

\title{Titan: Earth-like on the Outside, Ocean World on the Inside}

\correspondingauthor{Shannon M. MacKenzie}
\email{shannon.mackenzie@jhuapl.edu}

\author[0000-0002-1658-9687]{Shannon M. MacKenzie}
\affiliation{Johns Hopkins University Applied Physics Laboratory, 11100 Johns Hopkins Road, Laurel, MD 20723, USA }

\author{Samuel P.D. Birch}
\affiliation{Massachusetts Institute of Technology}

\author{Sarah H\"{o}rst}
\affiliation{Johns Hopkins University}

\author{Christophe Sotin}
\affiliation{Jet Propulsion Laboratory, California Institute of Technology}

\author{Erika Barth}
\affiliation{Southwest Research Institute, Boulder, Colorado, USA}

\author{Juan M. Lora}
\affiliation{Yale University}

\author{Melissa G. Trainer}
\affiliation{NASA Goddard Space Flight Center}

\author{Paul Corlies} 
\affiliation{Massachusetts Institute of Technology}

\author{Michael J. Malaska}
\affiliation{Jet Propulsion Laboratory, California Institute of Technology}

\author{Ella Sciamma-O’Brien}
\affiliation{NASA Ames Space Science and Astrobiology Division, Astrophysics Branch, USA}

\author{Alexander E. Thelen}
\affiliation{NASA Goddard Space Flight Center}

\author{Elizabeth Turtle}
\affiliation{Johns Hopkins University Applied Physics Laboratory, 11100 Johns Hopkins Road, Laurel, MD 20723, USA }

\author{Jani Radebaugh}
\affiliation{Department of Geological Sciences, Brigham Young University, S-389 ESC Provo, UT 84602, United States}

\author{Jennifer Hanley}
\affiliation{Lowell Observatory}

\author{Anezina Solomonidou}
\affiliation{California Institute of Technology, Pasadena, CA, USA}

\author{Claire Newman}
\affiliation{Aeolis Research, 333 N. Dobson Road, Unit 5, Chandler, AZ 85224, USA}

\author{Leonardo Regoli}
\affiliation{Johns Hopkins University Applied Physics Laboratory, 11100 Johns Hopkins Road, Laurel, MD 20723, USA }

\author{S\'{e}bastien Rodriguez}
\affiliation{Université de Paris, Institut de Physique du Globe de Paris, CNRS}

\author{Ben\^{o}it Seignovert}
\affiliation{Laboratoire de Planétologie et Géodynamique, Université de Nantes, Nantes, France}

\author{Alexander G. Hayes}
\affiliation{Cornell University, Ithaca NY, USA}

\author{Baptiste Journaux}
\affiliation{University of Washington}

\author{Jordan Steckloff}
\affiliation{Planetary Science Institute}

\author{Delphine Nna-Mvondo}
\affiliation{University of Maryland Baltimore County, Center for Space Sciences and Technology, Baltimore, Maryland, USA}

\author{Thomas Cornet}
\affiliation{European Space Agency (ESA), European Space Astronomy Centre (ESAC), Villanueva de la Canada, Madrid, Spain}

\author{Maureen Palmer}
\affiliation{Lunar and Planetary Laboratory, University of Arizona, 1629 E. University Blvd., Tucson, AZ 85721, USA}

\author{Rosaly M.C. Lopes}
\affiliation{Jet Propulsion Laboratory, California Institute of Technology.
}

\author{Sandrine Vinatier}
\affiliation{LESIA, Observatoire de Paris, Université PSL, CNRS, Sorbonne Université, Université de Paris, 5 place Jules Janssen, 92195 Meudon, France}

\author{Ralph Lorenz}
\affiliation{Johns Hopkins University Applied Physics Laboratory, 11100 Johns Hopkins Road, Laurel, MD 20723, USA }

\author{Conor Nixon}
\affiliation{NASA Goddard Space Flight Center}
\author{Ellen Czaplinski}
\affiliation{University of Arkansas}

\author{Jason W. Barnes}
\affiliation{Department of Physics, University of Idaho, Moscow, Idaho, USA}

\author{Ed Sittler}
\affiliation{NASA Goddard Space Flight Center}

\author{Andrew Coates}
\affiliation{Mullard Space Science Laboratory, University College London}

\begin{abstract}
Thanks to the \emph{Cassini-Huygens} mission, Titan, the pale orange dot of Pioneer and Voyager encounters has been revealed to be a dynamic, hydrologically-shaped, organic-rich ocean world offering unparalleled opportunities to explore prebiotic chemistry.  And while \emph{Cassini-Huygens} revolutionized our understanding of each of the three ``layers” of Titan—the atmosphere, the surface, and the interior—we are only beginning to hypothesize how these realms interact. In this paper, we summarize the current state of Titan knowledge and discuss how future exploration of Titan would address some of the next decade’s most compelling planetary science questions. We also demonstrate why exploring Titan, both with and beyond the \emph{Dragonfly} New Frontiers mission, is a necessary and complementary component of an Ocean Worlds Program that seeks to understand whether habitable environments exist elsewhere in our solar system. 
\end{abstract}


\section{Introduction} \label{sec:intro}
At the turn of the millennium, advancements in ground-based and space-based telescopes enabled the first glimpses of Titan's surface \citep[e.g.][]{smith_titans_1996,meier_surface_2000,coustenis_images_2001,coustenis_titans_2003,griffith_evidence_2003}. The heterogeneous surface albedo was inconsistent with a global ocean, immediately prompting discussion over the fate of ethane, anticipated to be one of the most abundant products of the atmospheric photochemistry \citep{lunine_ethane_1983}. When \emph{Cassini-Huygens} arrived in the Saturn system in 2004, a slew of fundamental advances were enabled by the combined in situ and remote sensing observations over the next 13 years. Newly identified features and processes prompted new questions, many of which will remain unanswered until a return mission to Titan \citep[e.g.][]{nixon_titans_2018,rodriguez_poseidon}. 

In light of these revelations--that Titan is a an organic-rich, ocean world where Earth-like geological processes rework the landscape and the complex atmospheric products that fall upon it--Titan was considered a target of high importance going into the 2012-2023 Decadal Survey \citep{national_research_council_vision_2011}. Similar to the Titan Explorer study of 2007 \citep{lockwood2008titan,lorenz2008titan}, the Titan Saturn System Mission concept study (TSSM) employed a comprehensive, three-pronged approach: an orbiter, a lander (targeting the lander to the northern lakes rather than the equatorial dunes), and a montgolfiere \citep{coustenis2009joint,reh2009titan}. This particular mission architecture was ultimately not put forward as the highest priority, instead deferred to the next decade ``primarily because of the greater technical readiness of [the Europa flagship mission]". The Vision and Voyages report further noted that ``[a Titan-returning mission's] high scientific priority, however, is especially noteworthy" and thus recommended continued development of the technologies needed to support such a mission.

As the 2012-2023 decade unfolded, however, new technologies, scientific revelations, and congressional inertia motivated the addition of Titan and Enceladus to the New Frontiers 4 competition. With the selection of \emph{Dragonfly}, some--but not all--of the high priority science identified by the TSSM will be addressed. Where, then, does Titan science stand now and what questions will be beyond the scope of \emph{Dragonfly}? How is the exploration of Titan's atmospheric, surface, and subsurface processes relevant to Ocean Worlds and other planets? We discuss the answers to these questions as the community enters the purview of a new decadal survey, stemming from considerations submitted as a white paper by these same authors to the 2023-2032 National Academies Planetary Science and Astrobiology Decadal Survey.

\section{Titan is an Organic World}
\begin{figure}
    \centering
    \includegraphics[width=0.7\textwidth]{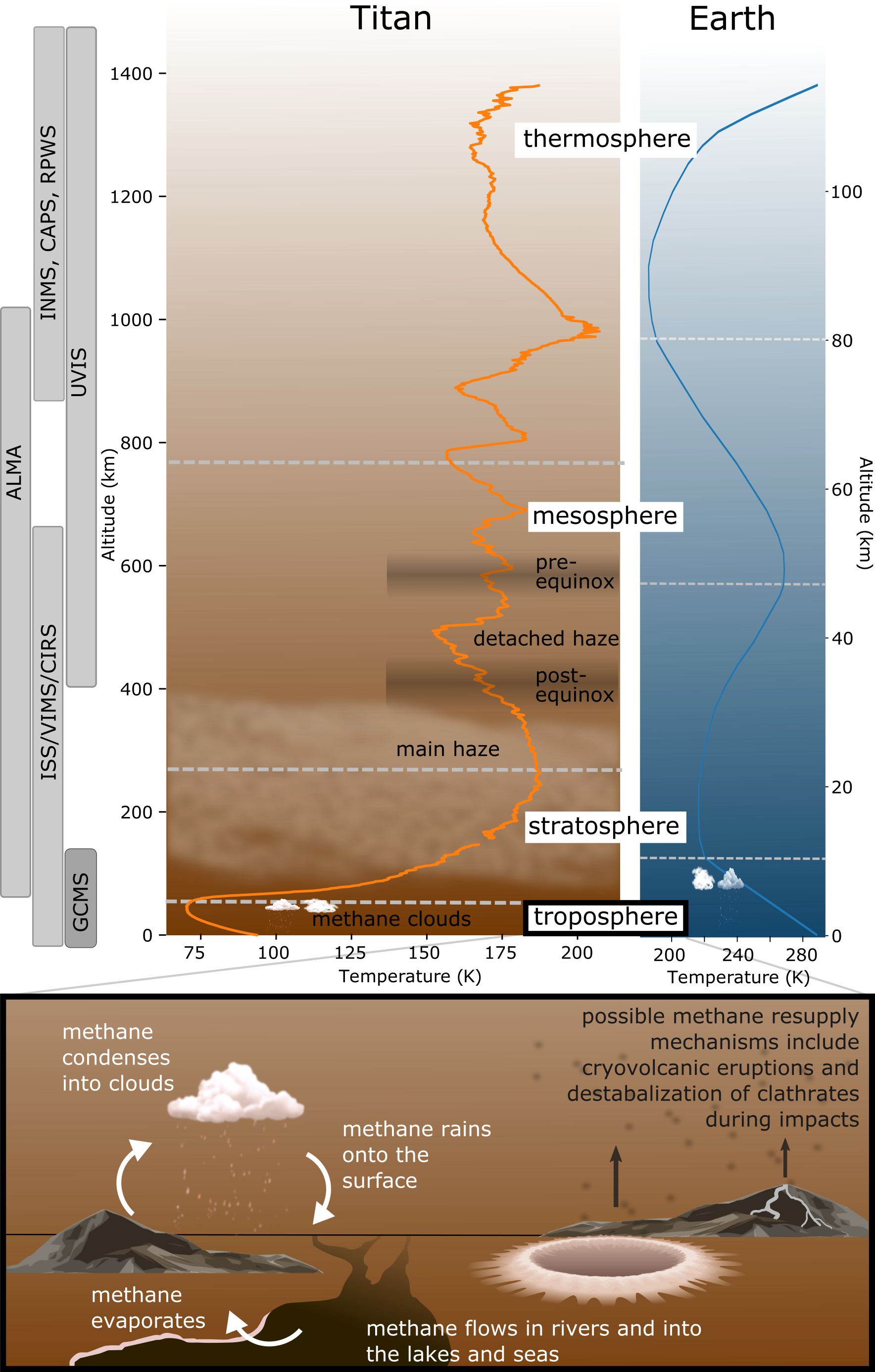}
    \caption{(top) Comparison of the temperature profiles and structure of the atmospheres of Titan and Earth. Vertical bars at left show the regions probed by Cassini instruments (Ion and Neutral Mass Spectrometer, INMS; CAssini Plasma Spectrometer, CAPS; Radio and Plasma Wave Science, RPWS; UltraViolet Imaging Spectrometer, UVIS; Imaging Science Subsystem, ISS; Visual and Infrared Mapping Spectrometer, VIMS; Composite InfraRed Spectrometer, CIRS), the Huygens Gas Chromatograph Mass Spectrometer (GCMS), and the Atacama Large Millimeter/submillimeter Array (ALMA) telescope. (bottom) Similar to the cycling of water on Earth, Titan's troposphere hosts the hydrological cycle of methane between the surface and on atmosphere. Irreversible loss of methane in the thermosphere suggests that a methane replenishment mechanism is necessary.}
    \label{atmofig}
\end{figure}

Titan hosts the most Earth-like atmosphere in the solar system. Similarities include the atmospheric structure \citep{fulchignoni_situ_2005} (Figure \ref{atmofig}), a nitrogen-dominated composition (95\% N$_2$, 4\% CH$_4$, and 1\% trace species at the surface), and a surface pressure of 1.5 bar. The photolytic destruction of atmospheric methane initiates a chain of photochemical reactions responsible for the plethora of organic species that make up the haze observed by \emph{Cassini-Huygens} and ground-based facilities \citep{marten_new_2002,gurwell_submillimeter_2004,ali_cyclopropenyl_2013,cordiner_alma_2014,cordiner_ethyl_2015,cordiner_interferometric_2018,cordier_floatability_2019,molter_alma_2016,desai_carbon_2017,lai_mapping_2017,palmer_alma_2017,teanby_origin_2018,lombardo_detection_2019,thelen_abundance_2019,thelen_measurement_2019,thelen_detection_2020,nixon_detection_2020}.  While unlike present-day Earth, Titan's haze production may be similar to processes on Early Earth \citep{trainer_organic_2006,g_trainer_atmospheric_2013}. 

\subsection{Pressing Questions and Future Investigations}
\emph{Cassini} identified the compositions of neutral and positive ions up to 99 Da but could only detect the presence of negative ions with a mass-to-charge ratio so large as to be similar to that of terrestrial proteins \citep{coates_discovery_2007,waite_process_2007,wellbrock_cassini_2013,woodson_ion_2015}. These molecules grow larger during polar winter and with decreasing altitude \citep{wellbrock_heavy_2019}, but their ultimate fate remains unknown. Thus, our current list of known compounds represents only the tip of the organic factory iceberg. Ongoing modeling efforts and laboratory experiments continue to investigate what reactions might be at work \citep{rannou_coupled_2004,waite_process_2007,vuitton_ion_2007,vuitton_simulating_2019,krasnopolsky_photochemical_2009,krasnopolsky_chemical_2014,nixon_isotopic_2012,waite_model_2013,larson_simulating_2014,dobrijevic_coupling_2014,dobrijevic_1d-coupled_2016,hickson_evolution_2014,lara_time-dependent_2014,loison_neutral_2015,loison_photochemical_2019,luspay-kuti_effects_2015,wong_plutos_2015,sebree_13c_2016,barth_modeling_2017,douglas_low_2018,mukundan_model_2018,berry_chemical_2019,bourgalais_low-pressure_2019,dubois_nitrogen-containing_2019,dubois_situ_2019,dubois_positive_2020} and to explore the properties of Titan haze analogs, tholins \citep{cable_titan_2012,gautier_influence_2017,sciamma-obrien_titan_2017,horst_haze_2018,sebree_detection_2018}. However, determining the dominant chemical pathways of Titan’s organic factory will require a more complete understanding of the chemical composition, abundance, and distribution of organic hazes in the atmosphere. Observations at different seasons in Titan’s year (29.5 Earth years) are critical, as monitoring with Cassini spanned only from the middle of northern winter to the very early summer \citep[e.g.][]{teanby_seasonal_2019,vinatier_optical_2012,vinatier_temperature_2020,seignovert_haze_2021}. Global atmospheric circulation in the upper atmosphere, which shows strong zonal prograde winds rapidly changing close to the northern summer solstice, has only been investigated very recently with ALMA \citep{lellouch_intense_2019,cordiner_detection_2020}. These wind changes also propagate in the middle atmosphere, as evidenced by CIRS data in the same period \citep{vinatier_temperature_2020,sharkey_potential_2021}. Observations from an orbiter would be necessary to monitor interactions between the middle and upper atmospheric circulation as well as their potential couplings with atmospheric chemistry--all of which are currently poorly known.

Beyond chemical pathways, many questions remain or have arisen from \emph{Cassini-Huygens}. How seasonal trends in condensate distributions extend to lower altitudes \citep{coates_discovery_2007,desai_carbon_2017} and whether these affect sedimentation onto the surface remains unknown. Models indicate that microphysics plays an important role in cloud and haze formation \citep{barth_modeling_2017}; increased understanding of the distribution, optical properties, and composition of hazes and clouds from observational (at Titan and ground-based) \citep{jennings_first_2012,jennings_seasonal_2012,jennings_evolution_2015,vinatier_optical_2012,seignovert_aerosols_2017,seignovert_haze_2021,le_mouelic_mapping_2018,west_seasonal_2018,anderson_organic_2018} and laboratory data \citep{anderson_spectral_2018,nna-mvondo_detailed_2019} would constrain physical models \citep[e.g.][]{loison_gas-grain_2020}.
Several approaches to constrain the age of Titan’s atmosphere overlap at $\sim$300-500 Myr, but whether this indicates the age of the atmosphere or ongoing methane photolysis is unclear \citep{horst_titans_2017}. Replenishment from the interior offers a compelling solution to the methane loss, but more data are required to evaluate the likelihood of candidate mechanisms like clathrate dissociation and cryovolcanism \citep{lunine_clathrate_1987,tobie_episodic_2006,choukroun_stability_2010,choukroun_is_2012,sotin_observations_2012}. Examples of necessary data include crustal dynamics, surface feature identification, isotopic fractionation, and a high-degree gravity field.

\subsection{Role in an Ocean Worlds Program} 
Atmospheric organic species are the ultimate source of the organic sediments that dominate Titan’s surface, so their formation and evolution in the atmosphere have important implications for surface geology and possible subsurface nutrient availability. Furthermore, investigating the processes that create complex species in Titan’s atmosphere—without, presumably, biological catalysts like those responsible for large molecules here on Earth—offers fundamental insight into the chemistry that may precede or facilitate the rise of biochemistry on Early Earth \citep{trainer_organic_2006,g_trainer_atmospheric_2013} and beyond. The study of Titan’s atmospheric chemistry therefore offers crucial context for the habitability potential of other ocean worlds where the essential elements may be less abundant. 

\subsection{Relevance to other planets}
\begin{figure}
    \centering
    \includegraphics[width=0.9\textwidth]{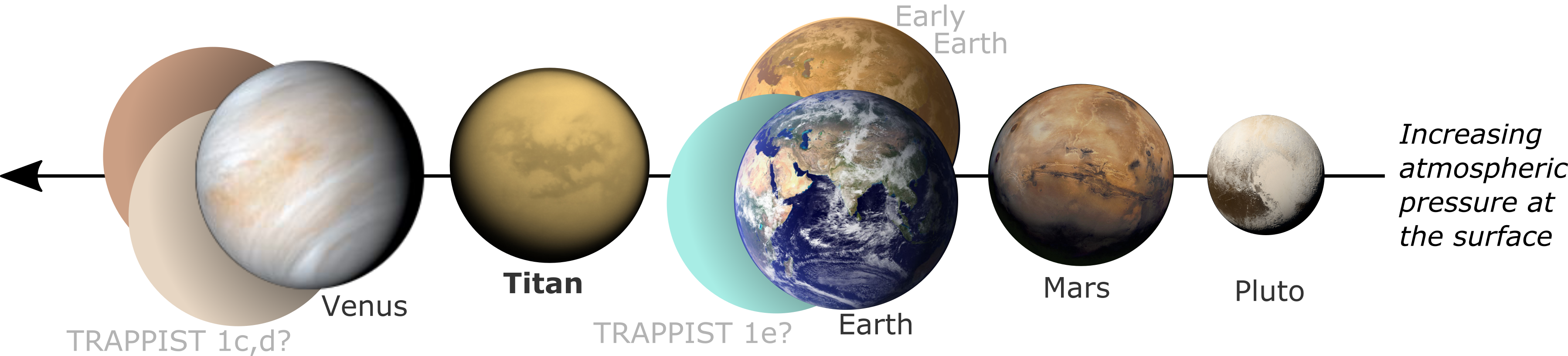}
    \caption{Titan's atmospheric density compared to that of terrestrial worlds both within and beyond \citep[e.g.][]{morley2017observing} our solar system, highlighting the potential for comparative planetology between Titan and other worlds with atmospheres.}
    \label{comparisons}
\end{figure}

Questions surrounding the dynamics and longevity of Titan’s atmosphere link to questions about the gas and ice giants \citep{robinson_titan_2014,toledo_constraints_2019} and—given the coupling between the atmosphere and surface—about Earth, Venus, Mars, and Pluto \citep{mitchell_effects_2014,mandt_comparative_2015,brain_atmospheric_2016,guendelman_axisymmetric_2018,read_superrotation_2018,crismani_localized_2019,kohn_streamer_2019,faulk_titans_2020,kite_methane_2020} (Figure \ref{comparisons}). Without its own magnetic field, Titan’s interactions with the solar wind and Saturn’s magnetosphere offer the opportunity to explore whether magnetic fields are necessary for habitability. Moreover, Titan’s atmosphere serves as a powerful backyard analog for hazy exoplanets—from understanding the formation and evolution of atmospheric aerosols to how we might best detect and observe them—as we have ground truth from both remote and in situ sensing \citep{de_kok_influence_2012,forget_possible_2014,tokano_precipitation_2015,arney_pale_2016,checlair_titan-like_2016,munoz_titan_2017,he_carbon_2017,horst_haze_2018,levi_equation_2019,lora_atmospheric_2018,alvarez_navarro_effects_2019,martinez-rodriguez_exomoons_2019,miguel_observability_2019}.

\section{Titan is an Active Hydrological and Sedimentary World}
\begin{figure}
    \centering
    \includegraphics[width=0.65\textwidth]{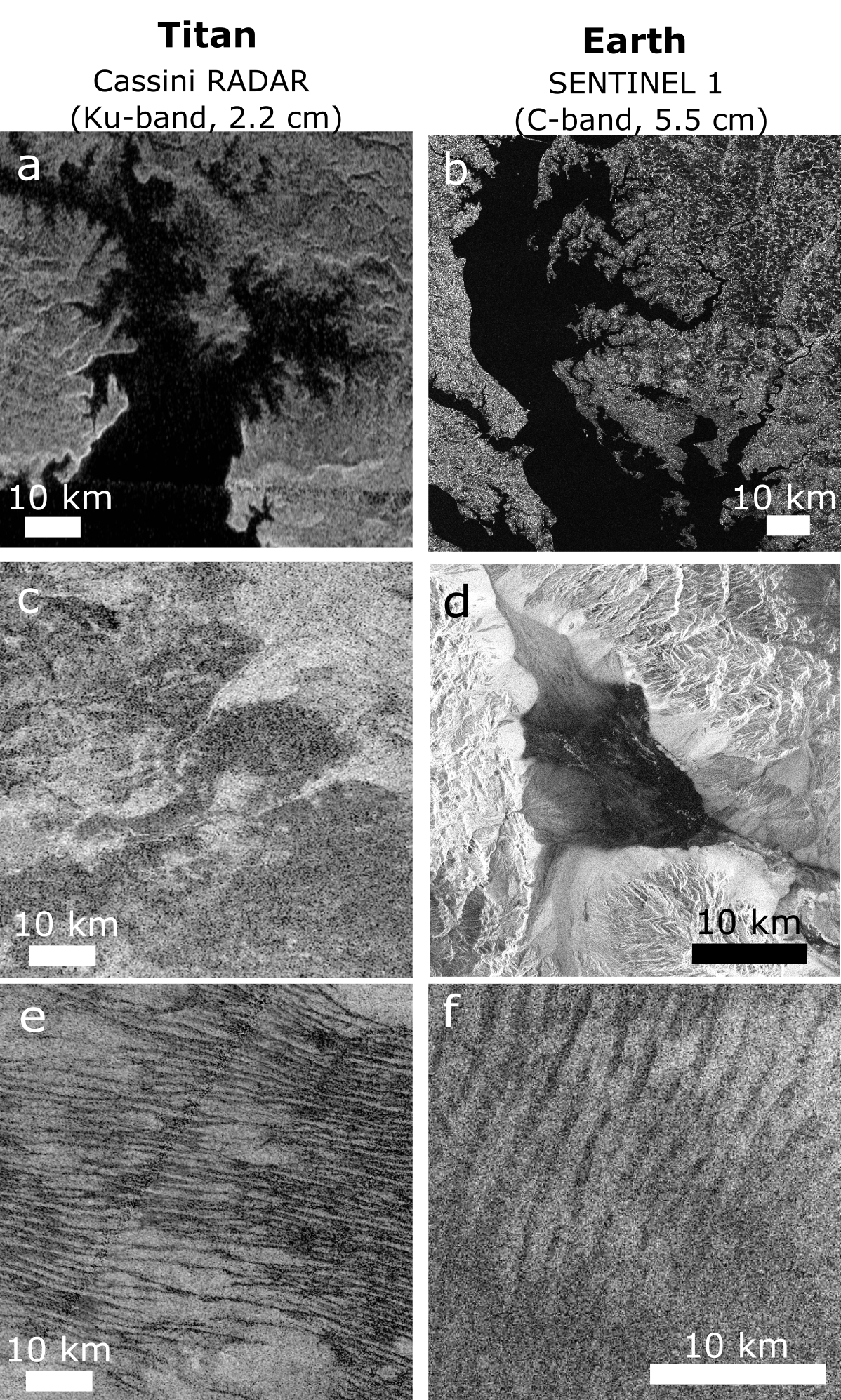}
    \caption{Select examples of terrains shaped by Titan's hydrological and sedimentological processes as viewed by Cassini RADAR and terrestrial analogs observed by SENTINEL 1: (a) shorelines of Kraken Mare from T28; (b) Chesapeake Bay (38.884$^{\circ}$N, 76.398$^{\circ}$W) ; (c) river channels terminating in fans from Ta; (d) Death Valley channels and fans (36.688$^{\circ}$N,117.177$^{\circ}$W);
    (e) organic sands organized into dunes from T49; (f) Namibian longitudinal dunes (24.285$^{\circ}$S,15.437$^{\circ}$E).}
    \label{terrains}
\end{figure}

Titan also has a methane-based hydrologic cycle akin to Earth’s \citep{mitchell_climate_2016,hayes_post-cassini_2018}. The seasonal timing and magnitude of surface-atmosphere fluxes of methane, including observed clouds and rainstorms \citep{turtle_cassini_2009,turtle_rapid_2011,turtle_seasonal_2011,turtle_titans_2018,brown_clouds_2010,barnes_transmission_2013,lemmon_large-scale_2019,dhingra_observational_2019}, are probably linked to existing surface and subsurface liquid reservoirs \citep{mitchell_drying_2008,mitchell_climate_2016,tokano_limnological_2009,tokano_modeling_2019,tokano_stable_2020,lora_titans_2015,lora_gcm_2015,lora_model_2019,newman_simulating_2016,faulk_regional_2017,faulk_titans_2020}. The hydrological cycle shapes the surface, producing landforms that bear a striking resemblance to those found on Earth \citep{hayes_lakes_2016} (Figure \ref{terrains}). Lakes and seas of liquid methane and ethane \citep{brown_identification_2008,mastrogiuseppe_bathymetry_2014,mastrogiuseppe_radar_2016,mastrogiuseppe_bathymetry_2018} up to hundreds of meters deep \citep{mastrogiuseppe_bathymetry_2014,mastrogiuseppe_radar_2016,mastrogiuseppe_bathymetry_2018,stofan_lakes_2007} are found across Titan’s polar regions \citep{turtle_rapid_2011,sotin_observations_2012,barnes_transmission_2013,dhingra_observational_2019,barnes_production_2015,hofgartner_transient_2014,hofgartner_titans_2016,cornet_dissolution_2015,mackenzie_case_2019,solomonidou_spectral_2020}. River channels and rounded cobbles imaged by Huygens \citep{karkoschka_disr_2016} and the radar-bright channels \citep{barnes_global-scale_2007,lorenz_titans_2008,burr_fluvial_2009,burr_morphology_2013,le_gall_radar-bright_2010,cartwright_channel_2011,black_estimating_2012,langhans_titans_2013} and fans \citep{birch_alluvial_2016,radebaugh_alluvial_2018,cartwright_using_2017} observed by \emph{Cassini} \citep{wasiak_geological_2013,poggiali_liquid-filled_2016} demonstrate that Titan’s hydrologic cycle is intimately connected with the sedimentary cycle: complex organic compounds synthesized and advected in and by the atmosphere are further transported and modified across the surface by the only known active extraterrestrial hydrologic cycle.
Perhaps the best studied sediments on Titan are the organic sands that occupy 17\% of the moon’s surface \citep{soderblom_correlations_2007,barnes_spectroscopy_2008,bonnefoy_compositional_2016,le_gall_cassini_2011,le_gall_modeling_2014,rodriguez_global_2014,brossier_geological_2018}. Linear dunes (100s of kilometers long and $\sim$100 m in height) demonstrate the importance of aeolian processes and underlying topography in the redistribution of Titan’s organics \citep{rodriguez_global_2014,radebaugh_dunes_2008,radebaugh_linear_2010,lorenz_global_2009,malaska_geomorphological_2016,paillou_radar_2016,telfer_long-wavelength_2019}. Titan’s vast mid-latitude plains ($\sim$65\% of the surface) are also hypothesized to consist of organic materials \citep{malaska_material_2016,lopes_nature_2016,lopes_global_2020,solomonidou_spectral_2018,mackenzie_thermal_2019}, but their composition and origin remains unknown. 

\subsection{Pressing Questions and Future Investigations} Despite advances in our understanding of the landscapes of Titan, the composition of the surface remains ill-constrained. \emph{Cassini} observations, limited in spectral and spatial resolution, suggest two general categories of materials: organic-rich and water-ice rich \citep{barnes_spectroscopy_2008,rodriguez_global_2014,brossier_geological_2018,solomonidou_spectral_2018,soderblom_geology_2009,griffith_corridor_2019}. Continued laboratory and theoretical work into the possible compositions and physical properties of Titan's solid \citep{mendez_harper_electrification_2017,cable_acetylene-ammonia_2018,cable_co-crystal_2019,cable_properties_2020,maynard-casely_prospects_2018,yu_where_2018,yu_single_2020} and liquid \citep[e.g.][]{farnsworth_nitrogen_2019,hanley_effects_2020,engle_phase_2020,steckloff_stratification_2020,vu_rapid_2020} surface materials are crucial for informing interpretations of \emph{Cassini-Huygens} data and supporting future exploration of the surface like the \emph{Dragonfly} mission. 

Dramatic advances in our understanding of Titan’s seasonally-evolving weather and climate \citep{mitchell_dynamics_2006,mitchell_climate_2016,hayes_post-cassini_2018} are similarly accompanied by new questions and key unknowns. The principal mechanisms controlling the timing and distribution of humidity \citep{adamkovics_meridional_2016,adamkovics_observations_2017}, convection, methane cloud formation, and precipitation remain incompletely understood, as do the sources and sinks of atmospheric methane and the roles of atmospheric variability \citep{griffith_titans_2008,mitchell_impact_2009,roe_titans_2012,mitchell_climate_2016,hayes_post-cassini_2018} and transient phenomena like dust storms \citep{rodriguez_observational_2018} in the climate. Likewise, the impact of heterogeneous surface-atmosphere coupling—for example, how Titan’s lakes affect the north polar environment \citep{rafkin_air-sea_2020}—and the magnitude and importance of regional climate variability are still largely unexplored. Further observations of Titan’s weather phenomena, coupled with improvements in physical modeling, are needed to continue elucidating Titan’s climate system and to link synoptic-scale processes to those at global and interannual scales, as well as to their impacts on the surface.

Global, high-resolution imaging ($<$100m) would revolutionize our understanding of Titan’s landforms, how they interact with each other and the atmosphere, and how they have evolved in the same way that Mars Global Surveyor’s orbital campaign fundamentally changed the study of Mars. Similarly, the lack of knowledge of Titan's topography specifically limits study of the transport of liquids and sediments on the surface, as well as of the influence of the surface on the atmosphere. \emph{Cassini} data covers only 9\% of Titan at scales too coarse for detailed geophysical and hydrological analysis of hydrologic catchments, mountain wave effects, or orographic clouds and precipitation \citep{corlies_titans_2017}. \emph{Huygens} data offer higher resolution but only over a few square kilometers \citep{daudon_new_2020}. In conjunction with maps of surface composition at high spatial and spectral resolution, global imaging and topographic data would address fundamental questions surrounding the hydrological, sedimentological, and meteorological cycles of Titan, augmenting \emph{Cassini} data and complementing \emph{Dragonfly}’s planned local in situ investigations. 

\subsection{Role in an Ocean Worlds Program} 
\begin{figure}
    \centering
    \includegraphics[width=0.2\textwidth]{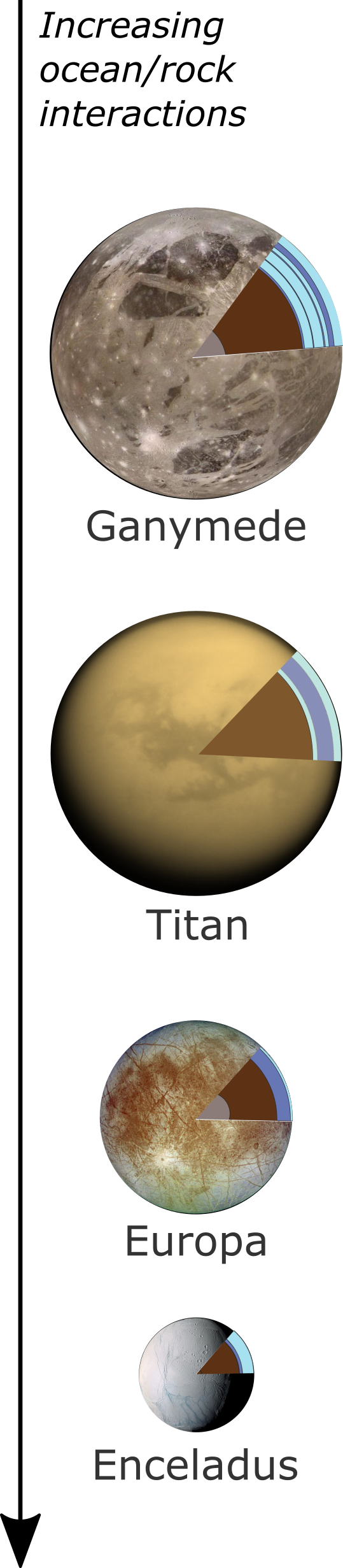}
    \caption{Titan on the water-rock interaction spectrum of Ocean Worlds, as anticipated from models of interior structure \citep[insets, based on the work of][]{vance_geophysical_2018}).}
    \label{comparisonsocean}
\end{figure}

Titan represents the organic-rich endmember of the Ocean World spectrum (Figure \ref{comparisonsocean}). Understanding the surface and atmospheric processes that create, modify, and transport these materials on Titan, and the timescales and volumes on which they act, would elucidate the role these processes play in planetary habitability and their significance. 

\subsection{Relevance to other planets}
Titan’s surface and climate system serves as a natural laboratory for studying the fundamentals of a planetary-scale hydrologic cycle, offering the unique opportunity to observe how this cycle controls the physical and chemical evolution of the landscape in an environment akin to but less complex than Earth’s. For example, sea level rise is likely ongoing and has dramatically shaped the coasts of Titan’s large seas \citep{aharonson_asymmetric_2009,hayes_transient_2011,hayes_topographic_2017,lora_simulations_2014,mackenzie_evidence_2014,mackenzie_case_2019,birch_morphological_2018,tokano_modeling_2019,tokano_stable_2020} and is likely ongoing although the rates remain loosely constrained; study of Titan’s coasts and ongoing erosional/depositional processes could be directly compared to the rapid changes on Earth and inform the study of paleo coastlines on Mars.

\section{Titan is an Ocean World}
A subsurface water ocean lies beneath Titan’s organic-covered ice crust \citep{nimmo_ocean_2016}, evidence for which includes gravitational tides \citep{iess_tides_2012,mitri_shape_2014} and larger-than-expected obliquity \citep{baland_titans_2011,baland_titans_2014}.  

\subsection{Pressing Questions and Future Investigations} 
The thickness of Titan’s crust is loosely constrained to 50-200 km \citep{choukroun_is_2012,nimmo_ocean_2016,hemingway_rigid_2013,lefevre_structure_2014} and the extent and duration of convection within the ice crust \citep{hemingway_rigid_2013,lefevre_structure_2014,noguchi_rheological_2020} is still debated. Estimates of the oceanic depth span 500-700 km \citep{iess_tides_2012,castillorogez_evolution_2010,gao_nonhydrostatic_2013,chen_tidal_2014} and the state of differentiation in the core is unknown \citep{nimmo_ocean_2016,baland_titans_2014,gao_nonhydrostatic_2013,orourke_stability_2014}. The presence of salts or ammonia may explain the ocean’s high density \citep{mitri_shape_2014,leitner_modeling_2019}, but magnesium sulfate is also a potential solution \citep{vance_geophysical_2018}. Primordial icy bodies provided noble gases and organic matter during Titan’s accretion, making the interior an even vaster source of organics than the atmosphere, with some models predicting 1000$\times$ the current atmospheric methane abundance \citep{tobie_titantextquotesingles_2012}. These considerations, coupled with detection of radiogenic $^{40}$Ar by \emph{Huygens} \citep{niemann_abundances_2005} suggest that outgassing from the interior may be responsible for the atmosphere. New isotopic measurements of noble gases and methane are necessary to resolve key questions concerning the ocean composition, the evolution of the interior and atmosphere, and the formation of Titan \citep{glein_noble_2015,glein_whiff_2017,marounina_evolution_2015,marounina_role_2018,miller_contributions_2019,journaux_holistic_2020}. 

At pressures $>$500 MPa, a layer of high-pressure ice may separate Titan’s core from the ocean \citep{vance_geophysical_2018}, but if the heat flux is high enough and/or salinity high enough, the ocean may be in direct contact with the silicate core \citep{journaux_large_2020}. Initially, the presence of high-pressure ice prompted the oceans of the largest icy satellites to be deemed inhospitable, assuming that separation by ice precluded exchange between the ocean and core. However, advances in our knowledge of how ices behave at high pressure show that convection can move material through the ice layer \citep{choblet_heat_2017,kalousova_two-phase_2018}, including salts and volatiles like $^{40}$Ar \citep{journaux_salt_2017,kalousova_melting_2018}. More laboratory and theoretical investigations into the properties of high pressure ices and hydrates are needed before we fully understand their implications on Ocean World habitability.

\subsection{Role in an Ocean Worlds Program} 
Determining whether Titan’s ocean is in contact with the rocky core would provide a key constraint to the formation and longevity of large Ocean Worlds both within and beyond our solar system \citep{journaux_influence_2013}. Studying the very origins of Titan’s organic cycle—from the primordial to hydrothermally altered material—informs our understanding of the role of volatile-rich ices in the early solar system. 

\section{Is Titan a Habitable World?}
The search for life elsewhere in the universe logically employs the guide of the biochemical foundations of Earth’s biosphere, giving rise to the classical conditions necessary for habitability: liquid water, essential elements (CHNO; the availabilities of P and S are yet to be determined), and energy sources \citep{hoehler_energy_2007,shock_quantitative_2007,domagal-goldman_astrobiology_2016}. All three factors exist on Titan. The question is where they have been or may be collocated and for how long. 

\subsection{Pressing Questions and Future Investigations} 
Titan’s deep crustal ice and subsurface ocean could be one of the largest habitable realms in the solar system, with a volume of liquid water 18x that of the Earth’s oceans and CHNOPS  \citep[potentially available from primordial and/or thermally processed materials;][]{miller_contributions_2019}. Tectonic activity and cryovolcanism may facilitate the delivery of surface organics through the crust. Whether any or all of these processes are at work and on what timescales they operate on Titan remain open questions, with implications for other ocean worlds where habitability may rely even more heavily upon the exchange of surface and subsurface material. For example, temperature and pressure conditions at the putative depth of Titan’s stagnant lid/convective ice transition are very similar to those encountered within terrestrial deep glacial ice, which hosts a diversity of microbial life \citep{miteva_comparison_2009} in the intergrain channels between solid ice grains \citep{price_microbial_2007,barletta_chemical_2012}. In these intergrain regions, microbial metabolism is slow enough that the environment may be habitable for 10,000 years—only a few orders of magnitude lower than Titan’s hypothesized convective cycle.  

A frigid ambient temperature of $\sim$90 K \citep{jennings_titans_2009,jennings_seasonal_2011,jennings_surface_2016,jennings_titan_2019,cottini_spatial_2012} makes Titan’s surface largely inhospitable for Earth-like life using water as the biochemical solvent. However, there are ephemeral scenarios in which liquid water is present at Titan’s surface: lavas erupting from cryovolcanoes and impact-generated melt. While some geomorphological evidence supports the existence of cryovolcanism \citep{lopes_global_2020,lopes_cryovolcanic_2007,lopes_cryovolcanism_2013}, its mechanics \citep{mitri_resurfacing_2008,moore_titan_2011} are not well understood, in part due to the lack of constraints on the extent, makeup, and activity of the crust as well as the ocean composition. However, impact craters are found across Titan’s surface \citep{lorenz_titans_2007,le_mouelic_mapping_2018,soderblom_geology_2010,neish_titans_2012,neish_crater_2013,neish_elevation_2014,neish_fluvial_2016,werynski_compositional_2019,hedgepeth_titans_2020,solomonidou_chemical_2020}. During the impact, crustal material and surface organics mix; the resulting pockets of liquid water eventually freeze on timescales loosely constrained to up to 10,000s of years \citep{artemieva_cratering_2003,artemieva_impact_2005,obrien_numerical_2005,neish_potential_2006,davies_atmospheric_2010,davies_cryolava_2016}. Mixing tholins with liquid water in the laboratory produces amino acids on a timescale of days \citep{neish_rate_2008,neish_low_2009,neish_titans_2010,neish_strategies_2018}. Titan’s transient liquid water environments are thus extraterrestrial laboratories for exploring how far prebiotic chemistry can progress under time and energy constraints that are difficult to realistically reproduce experimentally \citep{neish_strategies_2018}. The \emph{Dragonfly} mission will take advantage of this opportunity with surface composition measurements near a large impact crater. 

Without an understanding of the chemical processes necessary for the emergence of life, it is impossible to say with certainty how long it takes for life to arise \citep{orgel_origin_1998}. This timescale is a critical unknown in our concept of habitability: is there a minimum time necessary for all the key ingredients to be collocated? The answer to this question has immediate implications for strategizing the search for life elsewhere (both where to search and whether to target extant or extinct life), especially since the lifetime of the liquid oceans on both confirmed and candidate ocean worlds remains an active area of research \citep{nimmo_ocean_2016,neveu_evolution_2019}. Any constraints on habitability timescales from Titan’s transient liquid water environments would provide key context for exploration of potentially habitable environments and the search for life.

Finally, Titan’s lakes and seas of liquid hydrocarbons offer a unique opportunity to investigate whether the solvent necessary for biochemistry must be water. Theoretical considerations suggest alternative chemistries are possible \citep{benner_is_2004,lv_oxygen-free_2017} and the abundance of solid and liquid organic molecules available on the surface and lack of UV radiation make the surface of Titan an advantageous place for exploring the possibility of a true second genesis \citep{lunine_rivers_2009,lunine_titan_2010,mckay_titan_2016}. Theoretical investigations are exploring both the possibilities for lipid membrane-like structures in low temperature environments and whether cell membranes are even necessary \citep{palmer_alma_2017,stevenson_membrane_2015,rahm_polymorphism_2016,sandstrom_can_2020}. Laboratory and theoretical models are revolutionizing our understanding of the possible conditions within Titan’s lakes and seas \citep{cordier_floatability_2019,luspay-kuti_effects_2015,cordier_titans_2012,cordier_structure_2016,cordier_bubble_2017,cordier_bubbles_2018,hodyss_solubility_2013,corrales_acetonitrile_2017,malaska_laboratory_2017,hartwig_analytical_2018,czaplinski_experimental_2019,czaplinski_experimental_2020,farnsworth_nitrogen_2019}. Employing these new findings to constrain the habitability potential of Titan’s liquid hydrocarbons requires both determining the composition of Titan sediments—as the \emph{Dragonfly} mission’s plans to do by exploring at a portion of one of Titan's low-latitude dune fields—and monitoring the composition, physical conditions, and seasonal evolution of Titan’s polar lakes and seas with future missions. 

\section{Future Investigations at Titan}
\emph{Dragonfly}, the next New Frontiers (NF) mission, is a relocatable lander, to explore the prebiotic chemistry of Titan’s surface \citep{turtle_dragonfly_2017,lorenz_dragonfly_2018}. \citep[For a detailed description of \emph{Dragonfly}'s science goals and objectives, see][]{barnes2021dragonfly}.  Arriving in the 2030s, \emph{Dragonfly} will resolve a critical unknown: the chemical composition of Titan’s solid sediments. By using a mass spectrometer to measure compositions of the organic-rich sands of the equatorial dune fields, water-ice rich clasts from the relatively unaltered interdunes, and previously melted impact melt ejecta from an impact crater, \emph{Dragonfly} will begin to answer the question of how far prebiotic chemistry can progress in environments that provide long-term access to key ingredients for life, thereby providing crucial context for astrobiological investigations across the solar system. \emph{Dragonfly} will also determine elemental abundances in the near subsurface beneath the lander with a gamma ray neutron spectrometer, thus informing the availability and distribution of elements key to habitability.

Sample provenance both at the scale of \emph{Dragonfly}’s immediate environs and the local region is essential to interpreting the chemical findings in context. \emph{Dragonfly} is thus equipped with a suite of cameras to conduct imaging campaigns at local, nested scales. Meteorological and geophysical instruments will determine aeolian transport rates and monitor local weather conditions, as well as probing the thermal and electrical properties of the surface. Geophones and a seismometer round out the contextual measurements by probing the dynamics and properties of the ice crust, potentially constraining the depth to the ocean \citep{stahler_seismic_2018}. 

\emph{Dragonfly}’s payload is thus poised to revolutionize not only our understanding of Titan’s chemistry and geology but address more broadly how far prebiotic chemistry can progress and what chemical and geological processes make a planet or moon habitable. But, just as \emph{Curiosity} addresses different fundamental science than the \emph{Mars Reconnaissance Orbiter}, the NF-scope and architectural choices that make \emph{Dragonfly} best suited for its local in situ investigation necessarily preclude addressing many other outstanding questions at Titan, especially those requiring a global perspective. 

Thus, as demonstrated by exploration of Mars, a sequence of opportunities is needed to build upon and sufficiently leverage the detailed exploration of Titan begun by \emph{Cassini-Huygens} and to be continued by \emph{Dragonfly} in the coming decades. In particular, exploring the polar lakes and seas, their influence on Titan’s global hydrologic cycle, and their potential habitability, will remain out of even Dragonfly’s impressive range. Such measurements would also be complemented by orbital imaging at higher spatial and temporal resolutions than what \emph{Cassini} or ground-based observations could provide. A higher order gravity field would reveal eroded craters and thus constrain the prevalence of transient liquid water environments. More specifically, \emph{Dragonfly}’s seismic investigation of the interior would be significantly enhanced by a global topographic dataset and higher fidelity mapping of the gravity field. 

\begin{figure}
    \centering
    \includegraphics[width=0.85\textwidth]{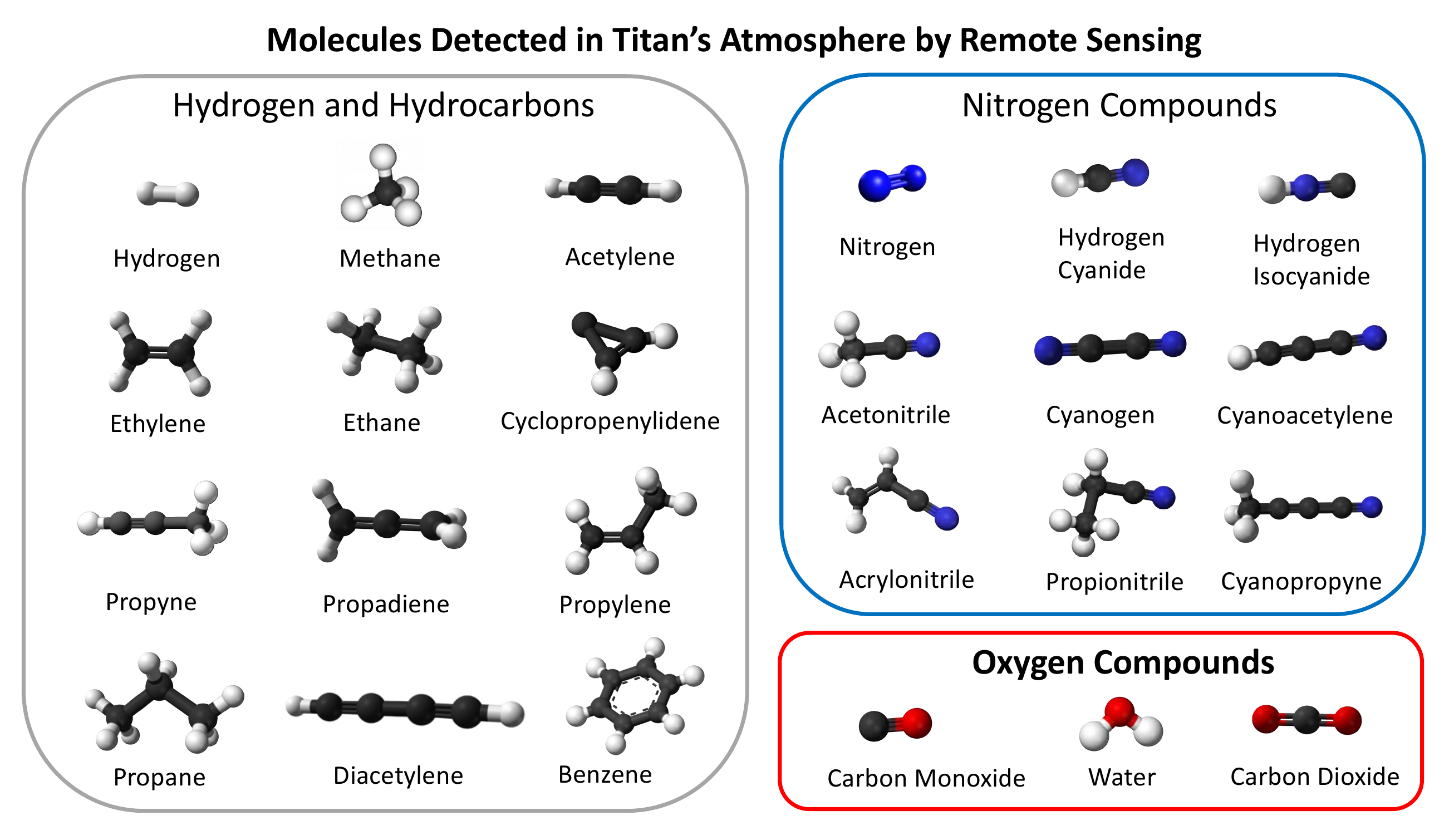}
    \caption{Molecules in Titan's atmosphere that have been uniquely identified via remote sensing. (Molecule image credit: Ben Mills/Wikimedia Commons) }
    \label{molecules}
\end{figure}

Further study of the dynamics of Titan’s climate and the seasonal evolution of hazes and weather phenomena \citep[e.g. clouds and haboobs,][]{smith_possible_2016,west_cassini_2016,le_mouelic_mapping_2018,rodriguez_observational_2018,stahler_seismic_2018,vinatier_study_2018,lemmon_large-scale_2019} requires continued long-term monitoring with ground- and space-based assets as Titan’s northern summer unfolds. A global imaging dataset would facilitate understanding the beginning-to-end life cycle of the materials sampled by \emph{Dragonfly}. Furthermore, as new species are identified in Titan's atmosphere, such as with ALMA (Figure \ref{molecules}), the needs of Titan exploration evolve. For example, as some of these species are only detected above 300 km and thus require orbital monitoring since low vapor pressures in the troposphere would make detection difficult for \emph{Dragonfly}. 

At least two examples for how to manifest these complementary investigations in the next decade were described in white papers to the 2023-2032 National Academies Planetary Science and Astrobiology Decadal Survey, representing New Frontiers and Flagship-scope efforts. But these are far from a comprehensive representation of possible architectures for returning to Titan.  A return to the Saturn system could orbit Titan \citep{sotin_oceanus_2017} for global mapping and geophysics or leverage the proximity of two prime Ocean World targets to jointly explore both Enceladus and Titan, via orbiting Saturn with plume flythroughs and frequent Titan flybys \citep{coustenis_tandem_2009,sotin_jet_2011} or shuttling between Titan and Enceladus \citep{russell_cycler_2009,sulaiman_joint}. Titan's thick atmosphere can be leveraged for long-duration flight \citep{lorenz_review_2008,barnes_aviatraerial_2012,ross_titan_2016} at altitudes high enough to maximize areal coverage and minimize atmospheric interference on compositional surface mapping \citep[e.g.][]{corlies_modeling_2021}. Ride-along small satellites can be exploited for gravity science \citep{tortora_ocean_2018}. On the surface, the diversity of interesting terrains inspired the study of a fleet of shape-changing robots \citep{tagliabue_shapeshifter_2020}, a fleet of mini-drones, and a drone capable of also floating on the surface of the seas \citep{rodriguez_poseidon}. A mission to float on Titan's seas has been proposed \citep{stofan_time_2013} and submerged instrumentation and/or vessels have been studied \citep[e.g.][]{lorenz_submarine_2016,lorenz_dropsonde_2018}. These in situ elements would benefit and/or require an orbiter for data relay. The diversity of mission concepts (and combinations thereof) that have been proposed and studied reflect the diversity of science questions left to answer at Titan and, importantly, demonstrate that compelling architectures span the full spectrum of NASA and ESA mission classes.

\section{Titan is an Unparalleled Destination}
Titan offers the opportunity to study a myriad of fundamental planetary science questions. The processes that govern its atmosphere, surface, and interior and interactions between these three environments make Titan an analog for destinations across the solar system and beyond. In the next decade, \emph{Dragonfly} will continue the legacy of \emph{Cassini-Huygens} and radically transform our understanding of Titan’s chemistry, geology, and astrobiological potential. But if the last decade has taught us anything, it’s that this moon’s complexity tends to defy our imagination. There is still much left to learn before we fully understand Saturn’s largest moon, requiring mission opportunities in addition to \emph{Dragonfly} in the next decade. 

\acknowledgements
We thank the broader Titan community for supporting the submission of a white paper of similar content to the 2023-2032 Decadal Survey. We also thank Caleb Heidel (JHU APL) for graphical contributions to Figure \ref{atmofig}. S.M.M. acknowledges support from JHU APL and Cassini Data Analysis and Participating Scientst Program (CDAP) grant \#80NSSC19K0888. S.P.D.B was supported by the Heising-Simons Foundation (51 Pegasi b Fellowship). Part of this work was conducted at the Jet Propulsion Laboratory, California Institute of Technology, under contract with NASA. R.L. and A.S. were partly supported by the CDAP grant \#NH16ZDA001N. This research was partly supported by the NASA Astrobiology Institute project entitled Habitability of Hydrocarbon Worlds: Titan and Beyond. D.N.-M gratefully acknowledges internal support received from the Center for Space Sciences and Technology of the University of Maryland Baltimore County.

\bibliography{bibliography}{}

\begin{thebibliography}{}
\expandafter\ifx\csname natexlab\endcsname\relax\def\natexlab#1{#1}\fi
\providecommand{\url}[1]{\href{#1}{#1}}
\providecommand{\dodoi}[1]{doi:~\href{http://doi.org/#1}{\nolinkurl{#1}}}
\providecommand{\doeprint}[1]{\href{http://ascl.net/#1}{\nolinkurl{http://ascl.net/#1}}}
\providecommand{\doarXiv}[1]{\href{https://arxiv.org/abs/#1}{\nolinkurl{https://arxiv.org/abs/#1}}}

\bibitem[{\'{A}d\'{a}mkovics \& de~Pater(2017)}]{adamkovics_observations_2017}
\'{A}d\'{a}mkovics, M., \& de~Pater, I. 2017, Icarus, 290, 134,
  \dodoi{10.1016/j.icarus.2017.02.015}

\bibitem[{\'{A}d\'{a}mkovics {et~al.}(2016)\'{A}d\'{a}mkovics, Mitchell, Hayes,
  Rojo, Corlies, Barnes, Ivanov, Brown, Baines, Buratti, Clark, Nicholson, \&
  Sotin}]{adamkovics_meridional_2016}
\'{A}d\'{a}mkovics, M., Mitchell, J.~L., Hayes, A.~G., {et~al.} 2016, Icarus,
  270, 376, \dodoi{10.1016/j.icarus.2015.05.023}

\bibitem[{Aharonson {et~al.}(2009)Aharonson, Hayes, Lunine, Lorenz, Allison, \&
  Elachi}]{aharonson_asymmetric_2009}
Aharonson, O., Hayes, A.~G., Lunine, J.~I., {et~al.} 2009, Nature Geoscience,
  2, 851, \dodoi{10.1038/ngeo698}

\bibitem[{Ali {et~al.}(2013)Ali, Sittler, Chornay, Rowe, \&
  Puzzarini}]{ali_cyclopropenyl_2013}
Ali, A., Sittler, E.~C., Chornay, D., Rowe, B.~R., \& Puzzarini, C. 2013,
  Planetary and Space Science, 87, 96, \dodoi{10.1016/j.pss.2013.07.007}

\bibitem[{Alvarez~Navarro {et~al.}(2019)Alvarez~Navarro, D\'{i}az,
  Garc\'{i}a-Ariza, \& Ram\'{i}rez}]{alvarez_navarro_effects_2019}
Alvarez~Navarro, E., D\'{i}az, B., Garc\'{i}a-Ariza, M.~A., \& Ram\'{i}rez,
  J.~E. 2019, The European Physical Journal Plus, 134, 458,
  \dodoi{10.1140/epjp/i2019-12826-4}

\bibitem[{Anderson {et~al.}(2018{\natexlab{a}})Anderson, Nna-Mvondo, Samuelson,
  McLain, \& Dworkin}]{anderson_spectral_2018}
Anderson, C.~M., Nna-Mvondo, D., Samuelson, R.~E., McLain, J.~L., \& Dworkin,
  J.~P. 2018{\natexlab{a}}, The Astrophysical Journal, 865, 62,
  \dodoi{10.3847/1538-4357/aadbab}

\bibitem[{Anderson {et~al.}(2018{\natexlab{b}})Anderson, Samuelson, \&
  Nna-Mvondo}]{anderson_organic_2018}
Anderson, C.~M., Samuelson, R.~E., \& Nna-Mvondo, D. 2018{\natexlab{b}}, Space
  Science Reviews, 214, 125, \dodoi{10.1007/s11214-018-0559-5}

\bibitem[{Arney {et~al.}(2016)Arney, Domagal-Goldman, Meadows, Wolf,
  Schwieterman, Charnay, Claire, Hébrard, \& Trainer}]{arney_pale_2016}
Arney, G., Domagal-Goldman, S.~D., Meadows, V.~S., {et~al.} 2016, Astrobiology,
  16, 873, \dodoi{10.1089/ast.2015.1422}

\bibitem[{Artemieva \& Lunine(2003)}]{artemieva_cratering_2003}
Artemieva, N., \& Lunine, J. 2003, Icarus, 164, 471,
  \dodoi{10.1016/S0019-1035(03)00148-9}

\bibitem[{Artemieva \& Lunine(2005)}]{artemieva_impact_2005}
Artemieva, N., \& Lunine, J.~I. 2005, Icarus, 175, 522,
  \dodoi{10.1016/j.icarus.2004.12.005}

\bibitem[{Baland {et~al.}(2011)Baland, Hoolst, Yseboodt, \&
  Karatekin}]{baland_titans_2011}
Baland, R.-M., Hoolst, T.~V., Yseboodt, M., \& Karatekin, O. 2011, Astronomy \&
  Astrophysics, 530, A141, \dodoi{10.1051/0004-6361/201116578}

\bibitem[{Baland {et~al.}(2014)Baland, Tobie, Lefèvre, \&
  Van~Hoolst}]{baland_titans_2014}
Baland, R.-M., Tobie, G., Lefèvre, A., \& Van~Hoolst, T. 2014, Icarus, 237,
  29, \dodoi{10.1016/j.icarus.2014.04.007}

\bibitem[{Barletta \& Roe(2012)}]{barletta_chemical_2012}
Barletta, R.~E., \& Roe, C.~H. 2012, Polar Record, 48, 334,
  \dodoi{10.1017/S0032247411000635}

\bibitem[{Barnes {et~al.}(2015)Barnes, Lorenz, Radebaugh, Hayes, Arnold, \&
  Chandler}]{barnes_production_2015}
Barnes, J.~W., Lorenz, R.~D., Radebaugh, J., {et~al.} 2015, Planetary Science,
  4, 1, \dodoi{10.1186/s13535-015-0004-y}

\bibitem[{Barnes {et~al.}(2007)Barnes, Brown, Soderblom, Buratti, Sotin,
  Rodriguez, Le~Mouèlic, Baines, Clark, \&
  Nicholson}]{barnes_global-scale_2007}
Barnes, J.~W., Brown, R.~H., Soderblom, L., {et~al.} 2007, Icarus, 186, 242,
  \dodoi{10.1016/j.icarus.2006.08.021}

\bibitem[{Barnes {et~al.}(2008)Barnes, Brown, Soderblom, Sotin, Le~Mouèlic,
  Rodriguez, Jaumann, Beyer, Buratti, Pitman, Baines, Clark, \&
  Nicholson}]{barnes_spectroscopy_2008}
---. 2008, Icarus, 195, 400, \dodoi{10.1016/j.icarus.2007.12.006}

\bibitem[{Barnes {et~al.}(2012)Barnes, Lemke, Foch, McKay, Beyer, Radebaugh,
  Atkinson, Lorenz, Le~Mouélic, Rodriguez, Gundlach, Giannini, Bain, Flasar,
  Hurford, Anderson, Merrison, \'{A}d\'{a}mkovics, Kattenhorn, Mitchell, Burr,
  Colaprete, Schaller, Friedson, Edgett, Coradini, Adriani, Sayanagi, Malaska,
  Morabito, \& Reh}]{barnes_aviatraerial_2012}
Barnes, J.~W., Lemke, L., Foch, R., {et~al.} 2012, Experimental Astronomy, 33,
  55, \dodoi{10.1007/s10686-011-9275-9}

\bibitem[{Barnes {et~al.}(2013)Barnes, Clark, Sotin, \'{A}d\'{a}mkovics,
  Appéré, Rodriguez, Soderblom, Brown, Buratti, Baines, Mouélic, \&
  Nicholson}]{barnes_transmission_2013}
Barnes, J.~W., Clark, R.~N., Sotin, C., {et~al.} 2013, The Astrophysical
  Journal, 777, 161, \dodoi{10.1088/0004-637X/777/2/161}

\bibitem[{Barnes(submitted)}]{barnes2021dragonfly}
Barnes, J. W. e.~a. submitted, Planetary Science Journal

\bibitem[{Barth(2017)}]{barth_modeling_2017}
Barth, E.~L. 2017, Planetary and Space Science, 137, 20,
  \dodoi{10.1016/j.pss.2017.01.003}

\bibitem[{Benner {et~al.}(2004)Benner, Ricardo, \& Carrigan}]{benner_is_2004}
Benner, S.~A., Ricardo, A., \& Carrigan, M.~A. 2004, Current Opinion in
  Chemical Biology, 8, 672, \dodoi{10.1016/j.cbpa.2004.10.003}

\bibitem[{Berry {et~al.}(2019)Berry, Ugelow, Tolbert, \&
  Browne}]{berry_chemical_2019}
Berry, J.~L., Ugelow, M.~S., Tolbert, M.~A., \& Browne, E.~C. 2019, ACS Earth
  and Space Chemistry, 3, 202, \dodoi{10.1021/acsearthspacechem.8b00139}

\bibitem[{Birch {et~al.}(2016)Birch, Hayes, Howard, Moore, \&
  Radebaugh}]{birch_alluvial_2016}
Birch, S. P.~D., Hayes, A.~G., Howard, A.~D., Moore, J.~M., \& Radebaugh, J.
  2016, Icarus, 270, 238, \dodoi{10.1016/j.icarus.2016.02.013}

\bibitem[{Birch {et~al.}(2018)Birch, Hayes, Corlies, Stofan, Hofgartner, Lopes,
  Lorenz, Lunine, MacKenzie, Malaska, Wood, \& Team}]{birch_morphological_2018}
Birch, S. P.~D., Hayes, A.~G., Corlies, P., {et~al.} 2018, Icarus, 310, 140,
  \dodoi{10.1016/j.icarus.2017.12.016}

\bibitem[{Black {et~al.}(2012)Black, Perron, Burr, \&
  Drummond}]{black_estimating_2012}
Black, B.~A., Perron, J.~T., Burr, D.~M., \& Drummond, S.~A. 2012, Journal of
  Geophysical Research: Planets, 117,
  \dodoi{https://doi.org/10.1029/2012JE004085}

\bibitem[{Bonnefoy {et~al.}(2016)Bonnefoy, Hayes, Hayne, Malaska, Le~Gall,
  Solomonidou, \& Lucas}]{bonnefoy_compositional_2016}
Bonnefoy, L.~E., Hayes, A.~G., Hayne, P.~O., {et~al.} 2016, Icarus, 270, 222,
  \dodoi{10.1016/j.icarus.2015.09.014}

\bibitem[{Bourgalais {et~al.}(2019)Bourgalais, Carrasco, Vettier, \&
  Pernot}]{bourgalais_low-pressure_2019}
Bourgalais, J., Carrasco, N., Vettier, L., \& Pernot, P. 2019, Journal of
  Geophysical Research: Space Physics, 124, 9214,
  \dodoi{https://doi.org/10.1029/2019JA026953}

\bibitem[{Brain {et~al.}(2016)Brain, Bagenal, Ma, Nilsson, \&
  Wieser}]{brain_atmospheric_2016}
Brain, D.~A., Bagenal, F., Ma, Y.-J., Nilsson, H., \& Wieser, G.~S. 2016,
  Journal of Geophysical Research: Planets, 121, 2364,
  \dodoi{https://doi.org/10.1002/2016JE005162}

\bibitem[{Brossier {et~al.}(2018)Brossier, Rodriguez, Cornet, Lucas, Radebaugh,
  Maltagliati, Mouélic, Solomonidou, Coustenis, Hirtzig, Jaumann, Stephan, \&
  Sotin}]{brossier_geological_2018}
Brossier, J.~F., Rodriguez, S., Cornet, T., {et~al.} 2018, Journal of
  Geophysical Research: Planets, 123, 1089,
  \dodoi{https://doi.org/10.1029/2017JE005399}

\bibitem[{Brown {et~al.}(2010)Brown, Roberts, \& Schaller}]{brown_clouds_2010}
Brown, M.~E., Roberts, J.~E., \& Schaller, E.~L. 2010, Icarus, 205, 571,
  \dodoi{10.1016/j.icarus.2009.08.024}

\bibitem[{Brown {et~al.}(2008)Brown, Soderblom, Soderblom, Clark, Jaumann,
  Barnes, Sotin, Buratti, Baines, \& Nicholson}]{brown_identification_2008}
Brown, R.~H., Soderblom, L.~A., Soderblom, J.~M., {et~al.} 2008, Nature, 454,
  607, \dodoi{10.1038/nature07100}

\bibitem[{Burr {et~al.}(2013)Burr, Drummond, Cartwright, Black, \&
  Perron}]{burr_morphology_2013}
Burr, D.~M., Drummond, S.~A., Cartwright, R., Black, B.~A., \& Perron, J.~T.
  2013, Icarus, 226, 742, \dodoi{10.1016/j.icarus.2013.06.016}

\bibitem[{Burr {et~al.}(2009)Burr, Jacobsen, Roth, Phillips, Mitchell, \&
  Viola}]{burr_fluvial_2009}
Burr, D.~M., Jacobsen, R.~E., Roth, D.~L., {et~al.} 2009, Geophysical Research
  Letters, 36, \dodoi{https://doi.org/10.1029/2009GL040909}

\bibitem[{Cable {et~al.}(2012)Cable, H\"{o}rst, Hodyss, Beauchamp, Smith, \&
  Willis}]{cable_titan_2012}
Cable, M.~L., H\"{o}rst, S.~M., Hodyss, R., {et~al.} 2012, Chemical Reviews,
  112, 1882, \dodoi{10.1021/cr200221x}

\bibitem[{Cable {et~al.}(2019)Cable, Vu, Malaska, Maynard-Casely, Choukroun, \&
  Hodyss}]{cable_co-crystal_2019}
Cable, M.~L., Vu, T.~H., Malaska, M.~J., {et~al.} 2019, ACS Earth and Space
  Chemistry, 3, 2808, \dodoi{10.1021/acsearthspacechem.9b00275}

\bibitem[{Cable {et~al.}(2020)Cable, Vu, Malaska, Maynard-Casely, Choukroun, \&
  Hodyss}]{cable_properties_2020}
---. 2020, ACS Earth and Space Chemistry, 4, 1375,
  \dodoi{10.1021/acsearthspacechem.0c00129}

\bibitem[{Cable {et~al.}(2018)Cable, Vu, Maynard-Casely, Choukroun, \&
  Hodyss}]{cable_acetylene-ammonia_2018}
Cable, M.~L., Vu, T.~H., Maynard-Casely, H.~E., Choukroun, M., \& Hodyss, R.
  2018, ACS Earth and Space Chemistry, 2, 366,
  \dodoi{10.1021/acsearthspacechem.7b00135}

\bibitem[{Cartwright {et~al.}(2011)Cartwright, Clayton, \&
  Kirk}]{cartwright_channel_2011}
Cartwright, R., Clayton, J.~A., \& Kirk, R.~L. 2011, Icarus, 214, 561,
  \dodoi{10.1016/j.icarus.2011.03.011}

\bibitem[{Cartwright \& Burr(2017)}]{cartwright_using_2017}
Cartwright, R.~J., \& Burr, D.~M. 2017, Icarus, 284, 183,
  \dodoi{10.1016/j.icarus.2016.11.013}

\bibitem[{Castillo‐Rogez \& Lunine(2010)}]{castillorogez_evolution_2010}
Castillo‐Rogez, J.~C., \& Lunine, J.~I. 2010, Geophysical Research Letters,
  37, \dodoi{https://doi.org/10.1029/2010GL044398}

\bibitem[{Checlair {et~al.}(2016)Checlair, McKay, \&
  Imanaka}]{checlair_titan-like_2016}
Checlair, J., McKay, C.~P., \& Imanaka, H. 2016, Planetary and Space Science,
  129, 1, \dodoi{10.1016/j.pss.2016.03.012}

\bibitem[{Chen {et~al.}(2014)Chen, Nimmo, \& Glatzmaier}]{chen_tidal_2014}
Chen, E. M.~A., Nimmo, F., \& Glatzmaier, G.~A. 2014, Icarus, 229, 11,
  \dodoi{10.1016/j.icarus.2013.10.024}

\bibitem[{Choblet {et~al.}(2017)Choblet, Tobie, Sotin, Kalousov\'{a}, \&
  Grasset}]{choblet_heat_2017}
Choblet, G., Tobie, G., Sotin, C., Kalousov\'{a}, K., \& Grasset, O. 2017,
  Icarus, 285, 252, \dodoi{10.1016/j.icarus.2016.12.002}

\bibitem[{Choukroun {et~al.}(2010)Choukroun, Grasset, Tobie, \&
  Sotin}]{choukroun_stability_2010}
Choukroun, M., Grasset, O., Tobie, G., \& Sotin, C. 2010, Icarus, 205, 581,
  \dodoi{10.1016/j.icarus.2009.08.011}

\bibitem[{Choukroun \& Sotin(2012)}]{choukroun_is_2012}
Choukroun, M., \& Sotin, C. 2012, Geophysical Research Letters, 39,
  \dodoi{https://doi.org/10.1029/2011GL050747}

\bibitem[{Coates {et~al.}(2007)Coates, Crary, Lewis, Young, Waite, \&
  Sittler}]{coates_discovery_2007}
Coates, A.~J., Crary, F.~J., Lewis, G.~R., {et~al.} 2007, Geophysical Research
  Letters, 34, \dodoi{https://doi.org/10.1029/2007GL030978}

\bibitem[{Cordier \& Carrasco(2019)}]{cordier_floatability_2019}
Cordier, D., \& Carrasco, N. 2019, Nature Geoscience, 12, 315,
  \dodoi{10.1038/s41561-019-0344-4}

\bibitem[{Cordier {et~al.}(2016)Cordier, Cornet, Barnes, MacKenzie, Le~Bahers,
  Nna-Mvondo, Rannou, \& Ferreira}]{cordier_structure_2016}
Cordier, D., Cornet, T., Barnes, J.~W., {et~al.} 2016, Icarus, 270, 41,
  \dodoi{10.1016/j.icarus.2015.12.034}

\bibitem[{Cordier {et~al.}(2017)Cordier, Garc\'{i}a-S\'{a}nchez,
  Justo-Garc\'{i}a, \& Liger-Belair}]{cordier_bubble_2017}
Cordier, D., Garc\'{i}a-S\'{a}nchez, F., Justo-Garc\'{i}a, D.~N., \&
  Liger-Belair, G. 2017, Nature Astronomy, 1, 1,
  \dodoi{10.1038/s41550-017-0102}

\bibitem[{Cordier \& Liger-Belair(2018)}]{cordier_bubbles_2018}
Cordier, D., \& Liger-Belair, G. 2018, The Astrophysical Journal, 859, 26,
  \dodoi{10.3847/1538-4357/aabc10}

\bibitem[{Cordier {et~al.}(2012)Cordier, Mousis, Lunine, Lebonnois, Rannou,
  Lavvas, Lobo, \& Ferreira}]{cordier_titans_2012}
Cordier, D., Mousis, O., Lunine, J.~I., {et~al.} 2012, Planetary and Space
  Science, 61, 99, \dodoi{10.1016/j.pss.2011.05.009}

\bibitem[{Cordiner {et~al.}(2020)Cordiner, Garcia-Berrios, Cosentino, Teanby,
  Newman, Nixon, Thelen, \& Charnley}]{cordiner_detection_2020}
Cordiner, M.~A., Garcia-Berrios, E., Cosentino, R.~G., {et~al.} 2020, The
  Astrophysical Journal, 904, L12, \dodoi{10.3847/2041-8213/abc688}

\bibitem[{Cordiner {et~al.}(2018)Cordiner, Nixon, Charnley, Teanby, Molter,
  Kisiel, \& Vuitton}]{cordiner_interferometric_2018}
Cordiner, M.~A., Nixon, C.~A., Charnley, S.~B., {et~al.} 2018, The
  Astrophysical Journal, 859, L15, \dodoi{10.3847/2041-8213/aac38d}

\bibitem[{Cordiner {et~al.}(2014)Cordiner, Nixon, Teanby, Irwin, Serigano,
  Charnley, Milam, Mumma, Lis, Villanueva, Paganini, Kuan, \&
  Remijan}]{cordiner_alma_2014}
Cordiner, M.~A., Nixon, C.~A., Teanby, N.~A., {et~al.} 2014, The Astrophysical
  Journal, 795, L30, \dodoi{10.1088/2041-8205/795/2/L30}

\bibitem[{Cordiner {et~al.}(2015)Cordiner, Palmer, Nixon, Irwin, Teanby,
  Charnley, Mumma, Kisiel, Serigano, Kuan, Chuang, \&
  Wang}]{cordiner_ethyl_2015}
Cordiner, M.~A., Palmer, M.~Y., Nixon, C.~A., {et~al.} 2015, The Astrophysical
  Journal, 800, L14, \dodoi{10.1088/2041-8205/800/1/L14}

\bibitem[{Corlies {et~al.}(2017)Corlies, Hayes, Birch, Lorenz, Stiles, Kirk,
  Poggiali, Zebker, \& Iess}]{corlies_titans_2017}
Corlies, P., Hayes, A.~G., Birch, S. P.~D., {et~al.} 2017, Geophysical Research
  Letters, 44, 11,754, \dodoi{https://doi.org/10.1002/2017GL075518}

\bibitem[{Corlies {et~al.}(2021)Corlies, McDonald, Hayes, Wray,
  \'{A}d\'{a}mkovics, Malaska, Cable, Hofgartner, H\"{o}rst, Liuzzo, Buffo,
  Lorenz, \& Turtle}]{corlies_modeling_2021}
Corlies, P., McDonald, G.~D., Hayes, A.~G., {et~al.} 2021, Icarus, 357, 114228,
  \dodoi{10.1016/j.icarus.2020.114228}

\bibitem[{Cornet {et~al.}(2015)Cornet, Cordier, Bahers, Bourgeois, Fleurant,
  Mouélic, \& Altobelli}]{cornet_dissolution_2015}
Cornet, T., Cordier, D., Bahers, T.~L., {et~al.} 2015, Journal of Geophysical
  Research: Planets, 120, 1044, \dodoi{https://doi.org/10.1002/2014JE004738}

\bibitem[{Corrales {et~al.}(2017)Corrales, Yi, Trumbo, Shalloway, Lunine, \&
  Usher}]{corrales_acetonitrile_2017}
Corrales, L.~R., Yi, T.~D., Trumbo, S.~K., {et~al.} 2017, The Journal of
  Chemical Physics, 146, 104308, \dodoi{10.1063/1.4978395}

\bibitem[{Cottini {et~al.}(2012)Cottini, Nixon, Jennings, de~Kok, Teanby,
  Irwin, \& Flasar}]{cottini_spatial_2012}
Cottini, V., Nixon, C.~A., Jennings, D.~E., {et~al.} 2012, Planetary and Space
  Science, 60, 62, \dodoi{10.1016/j.pss.2011.03.015}

\bibitem[{Coustenis {et~al.}(2009{\natexlab{a}})Coustenis, Lunine, Matson,
  Hansen, Reh, Beauchamp, Lebreton, \& Erd}]{coustenis2009joint}
Coustenis, A., Lunine, J., Matson, D., {et~al.} 2009{\natexlab{a}}, in Lunar
  and Planetary Science Conference, 1060

\bibitem[{Coustenis {et~al.}(2003)Coustenis, Salama, Schulz, Ott, Lellouch,
  Encrenaz, Gautier, \& Feuchtgruber}]{coustenis_titans_2003}
Coustenis, A., Salama, A., Schulz, B., {et~al.} 2003, Icarus, 161, 383,
  \dodoi{10.1016/S0019-1035(02)00028-3}

\bibitem[{Coustenis {et~al.}(2001)Coustenis, Gendron, Lai, Véran, Woillez,
  Combes, Vapillon, Fusco, Mugnier, \& Rannou}]{coustenis_images_2001}
Coustenis, A., Gendron, E., Lai, O., {et~al.} 2001, Icarus, 154, 501,
  \dodoi{10.1006/icar.2001.6643}

\bibitem[{Coustenis {et~al.}(2009{\natexlab{b}})Coustenis, Atreya, Balint,
  Brown, Dougherty, Ferri, Fulchignoni, Gautier, Gowen, Griffith, Gurvits,
  Jaumann, Langevin, Leese, Lunine, McKay, Moussas, Müller-Wodarg, Neubauer,
  Owen, Raulin, Sittler, Sohl, Sotin, Tobie, Tokano, Turtle, Wahlund, Waite,
  Baines, Blamont, Coates, Dandouras, Krimigis, Lellouch, Lorenz, Morse, Porco,
  Hirtzig, Saur, Spilker, Zarnecki, Choi, Achilleos, Amils, Annan, Atkinson,
  Bénilan, Bertucci, Bézard, Bjoraker, Blanc, Boireau, Bouman, Cabane,
  Capria, Chassefière, Coll, Combes, Cooper, Coradini, Crary, Cravens, Daglis,
  de~Angelis, de~Bergh, de~Pater, Dunford, Durry, Dutuit, Fairbrother, Flasar,
  Fortes, Frampton, Fujimoto, Galand, Grasset, Grott, Haltigin, Herique,
  Hersant, Hussmann, Ip, Johnson, Kallio, Kempf, Knapmeyer, Kofman, Koop,
  Kostiuk, Krupp, Küppers, Lammer, Lara, Lavvas, Le~Mouélic, Lebonnois,
  Ledvina, Li, Livengood, Lopes, Lopez-Moreno, Luz, Mahaffy, Mall,
  Martinez-Frias, Marty, McCord, Menor~Salvan, Milillo, Mitchell, Modolo,
  Mousis, Nakamura, Neish, Nixon, Nna~Mvondo, Orton, Paetzold, Pitman,
  Pogrebenko, Pollard, Prieto-Ballesteros, Rannou, Reh, Richter, Robb, Rodrigo,
  Rodriguez, Romani, Ruiz~Bermejo, Sarris, Schenk, Schmitt, Schmitz,
  Schulze-Makuch, Schwingenschuh, Selig, Sicardy, Soderblom, Spilker, Stam,
  Steele, Stephan, Strobel, Szego, Szopa, Thissen, Tomasko, Toublanc, Vali,
  Vardavas, Vuitton, West, Yelle, \& Young}]{coustenis_tandem_2009}
Coustenis, A., Atreya, S.~K., Balint, T., {et~al.} 2009{\natexlab{b}},
  Experimental Astronomy, 23, 893, \dodoi{10.1007/s10686-008-9103-z}

\bibitem[{Crismani {et~al.}(2019)Crismani, Deighan, Schneider, Plane, Withers,
  Halekas, Chaffin, \& Jain}]{crismani_localized_2019}
Crismani, M. M.~J., Deighan, J., Schneider, N.~M., {et~al.} 2019, Journal of
  Geophysical Research: Space Physics, 124, 4870,
  \dodoi{https://doi.org/10.1029/2018JA026251}

\bibitem[{Czaplinski {et~al.}(2020)Czaplinski, Yu, Dzurilla, \&
  Chevrier}]{czaplinski_experimental_2020}
Czaplinski, E., Yu, X., Dzurilla, K., \& Chevrier, V. 2020, The Planetary
  Science Journal, 1, 76, \dodoi{10.3847/PSJ/abbf57}

\bibitem[{Czaplinski {et~al.}(2019)Czaplinski, Gilbertson, Farnsworth, \&
  Chevrier}]{czaplinski_experimental_2019}
Czaplinski, E.~C., Gilbertson, W.~A., Farnsworth, K.~K., \& Chevrier, V.~F.
  2019, ACS Earth and Space Chemistry, 3, 2353,
  \dodoi{10.1021/acsearthspacechem.9b00204}

\bibitem[{Daudon {et~al.}(2020)Daudon, Lucas, Rodriguez, Jacquemoud,
  Escalante~López, Grieger, Howington‐Kraus, Karkoschka, Kirk, Perron,
  Soderblom, \& Costa}]{daudon_new_2020}
Daudon, C., Lucas, A., Rodriguez, S., {et~al.} 2020, Earth and Space Science,
  7, e2020EA001127, \dodoi{https://doi.org/10.1029/2020EA001127}

\bibitem[{Davies {et~al.}(2016)Davies, Sotin, Choukroun, Matson, \&
  Johnson}]{davies_cryolava_2016}
Davies, A.~G., Sotin, C., Choukroun, M., Matson, D.~L., \& Johnson, T.~V. 2016,
  Icarus, 274, 23, \dodoi{10.1016/j.icarus.2016.02.046}

\bibitem[{Davies {et~al.}(2010)Davies, Sotin, Matson, Castillo-Rogez, Johnson,
  Choukroun, \& Baines}]{davies_atmospheric_2010}
Davies, A.~G., Sotin, C., Matson, D.~L., {et~al.} 2010, Icarus, 208, 887,
  \dodoi{10.1016/j.icarus.2010.02.025}

\bibitem[{de~Kok \& Stam(2012)}]{de_kok_influence_2012}
de~Kok, R.~J., \& Stam, D.~M. 2012, Icarus, 221, 517,
  \dodoi{10.1016/j.icarus.2012.08.020}

\bibitem[{Desai {et~al.}(2017)Desai, Coates, Wellbrock, Vuitton, Crary,
  Gonz\'{a}lez-Caniulef, Shebanits, Jones, Lewis, Waite, Cordiner, Taylor,
  Kataria, Wahlund, Edberg, \& Sittler}]{desai_carbon_2017}
Desai, R.~T., Coates, A.~J., Wellbrock, A., {et~al.} 2017, The Astrophysical
  Journal, 844, L18, \dodoi{10.3847/2041-8213/aa7851}

\bibitem[{Dhingra {et~al.}(2019)Dhingra, Barnes, Brown, Burrati, Sotin,
  Nicholson, Baines, Clark, Soderblom, Jauman, Rodriguez, Mouélic, Turtle,
  Perry, Cottini, \& Jennings}]{dhingra_observational_2019}
Dhingra, R.~D., Barnes, J.~W., Brown, R.~H., {et~al.} 2019, Geophysical
  Research Letters, 46, 1205, \dodoi{https://doi.org/10.1029/2018GL080943}

\bibitem[{Dobrijevic {et~al.}(2014)Dobrijevic, Hébrard, Loison, \&
  Hickson}]{dobrijevic_coupling_2014}
Dobrijevic, M., Hébrard, E., Loison, J.~C., \& Hickson, K.~M. 2014, Icarus,
  228, 324, \dodoi{10.1016/j.icarus.2013.10.015}

\bibitem[{Dobrijevic {et~al.}(2016)Dobrijevic, Loison, Hickson, \&
  Gronoff}]{dobrijevic_1d-coupled_2016}
Dobrijevic, M., Loison, J.~C., Hickson, K.~M., \& Gronoff, G. 2016, Icarus,
  268, 313, \dodoi{10.1016/j.icarus.2015.12.045}

\bibitem[{Domagal-Goldman {et~al.}(2016)Domagal-Goldman, Wright,
  Domagal-Goldman, Wright, Adamala, Arina de~la Rubia, Bond, Dartnell, Goldman,
  Lynch, Naud, Paulino-Lima, Singer, Walther-Antonio, Abrevaya, Anderson,
  Arney, Atri, Azúa-Bustos, Bowman, Brazelton, Brennecka, Carns, Chopra,
  Colangelo-Lillis, Crockett, DeMarines, Frank, Frantz, de~la Fuente, Galante,
  Glass, Gleeson, Glein, Goldblatt, Horak, Horodyskyj, Kaçar, Kereszturi,
  Knowles, Mayeur, McGlynn, Miguel, Montgomery, Neish, Noack, Petryshyn,
  Rugheimer, Stüeken, Tamez-Hidalgo, Walker, \&
  Wong}]{domagal-goldman_astrobiology_2016}
Domagal-Goldman, S.~D., Wright, K.~E., Domagal-Goldman, S.~D., {et~al.} 2016,
  Astrobiology, 16, 561, \dodoi{10.1089/ast.2015.1460}

\bibitem[{Douglas {et~al.}(2018)Douglas, Blitz, Feng, Heard, Plane, Slater,
  Willacy, \& Seakins}]{douglas_low_2018}
Douglas, K., Blitz, M.~A., Feng, W., {et~al.} 2018, Icarus, 303, 10,
  \dodoi{10.1016/j.icarus.2017.12.023}

\bibitem[{Dubois {et~al.}(2019{\natexlab{a}})Dubois, Carrasco, Bourgalais,
  Vettier, Desai, Wellbrock, \& Coates}]{dubois_nitrogen-containing_2019}
Dubois, D., Carrasco, N., Bourgalais, J., {et~al.} 2019{\natexlab{a}}, The
  Astrophysical Journal, 872, L31, \dodoi{10.3847/2041-8213/ab05e5}

\bibitem[{Dubois {et~al.}(2020)Dubois, Carrasco, Jovanovic, Vettier, Gautier,
  \& Westlake}]{dubois_positive_2020}
Dubois, D., Carrasco, N., Jovanovic, L., {et~al.} 2020, Icarus, 338, 113437,
  \dodoi{10.1016/j.icarus.2019.113437}

\bibitem[{Dubois {et~al.}(2019{\natexlab{b}})Dubois, Carrasco, Petrucciani,
  Vettier, Tigrine, \& Pernot}]{dubois_situ_2019}
Dubois, D., Carrasco, N., Petrucciani, M., {et~al.} 2019{\natexlab{b}}, Icarus,
  317, 182, \dodoi{10.1016/j.icarus.2018.07.006}

\bibitem[{Engle {et~al.}(2020)Engle, Hanley, Thompson, Lindberg, Dustrud,
  Grundy, \& Tegler}]{engle_phase_2020}
Engle, A., Hanley, J., Thompson, G., {et~al.} 2020, Bulletin of the AAS, 52.
\newblock \url{https://baas.aas.org/pub/2020n6i408p02/release/1}

\bibitem[{Farnsworth {et~al.}(2019)Farnsworth, Chevrier, Steckloff, Laxton,
  Singh, Soto, \& Soderblom}]{farnsworth_nitrogen_2019}
Farnsworth, K.~K., Chevrier, V.~F., Steckloff, J.~K., {et~al.} 2019,
  Geophysical Research Letters, 46, 13658,
  \dodoi{https://doi.org/10.1029/2019GL084792}

\bibitem[{Faulk {et~al.}(2020)Faulk, Lora, Mitchell, \&
  Milly}]{faulk_titans_2020}
Faulk, S.~P., Lora, J.~M., Mitchell, J.~L., \& Milly, P. C.~D. 2020, Nature
  Astronomy, 4, 390, \dodoi{10.1038/s41550-019-0963-0}

\bibitem[{Faulk {et~al.}(2017)Faulk, Mitchell, Moon, \&
  Lora}]{faulk_regional_2017}
Faulk, S.~P., Mitchell, J.~L., Moon, S., \& Lora, J.~M. 2017, Nature
  Geoscience, 10, 827, \dodoi{10.1038/ngeo3043}

\bibitem[{Forget \& Leconte(2014)}]{forget_possible_2014}
Forget, F., \& Leconte, J. 2014, Philosophical Transactions of the Royal
  Society A: Mathematical, Physical and Engineering Sciences, 372, 20130084,
  \dodoi{10.1098/rsta.2013.0084}

\bibitem[{Fulchignoni {et~al.}(2005)Fulchignoni, Ferri, Angrilli, Ball,
  Bar-Nun, Barucci, Bettanini, Bianchini, Borucki, Colombatti, Coradini,
  Coustenis, Debei, Falkner, Fanti, Flamini, Gaborit, Grard, Hamelin, Harri,
  Hathi, Jernej, Leese, Lehto, Lion~Stoppato, López-Moreno, Mäkinen,
  McDonnell, McKay, Molina-Cuberos, Neubauer, Pirronello, Rodrigo, Saggin,
  Schwingenschuh, Seiff, Simões, Svedhem, Tokano, Towner, Trautner, Withers,
  \& Zarnecki}]{fulchignoni_situ_2005}
Fulchignoni, M., Ferri, F., Angrilli, F., {et~al.} 2005, Nature, 438, 785,
  \dodoi{10.1038/nature04314}

\bibitem[{Gao \& Stevenson(2013)}]{gao_nonhydrostatic_2013}
Gao, P., \& Stevenson, D.~J. 2013, Icarus, 226, 1185,
  \dodoi{10.1016/j.icarus.2013.07.034}

\bibitem[{Gautier {et~al.}(2017)Gautier, Sebree, Li, Pinnick, Grubisic,
  Loeffler, Getty, Trainer, \& Brinckerhoff}]{gautier_influence_2017}
Gautier, T., Sebree, J.~A., Li, X., {et~al.} 2017, Planetary and Space Science,
  140, 27, \dodoi{10.1016/j.pss.2017.03.012}

\bibitem[{Glein(2015)}]{glein_noble_2015}
Glein, C.~R. 2015, Icarus, 250, 570, \dodoi{10.1016/j.icarus.2015.01.001}

\bibitem[{Glein(2017)}]{glein_whiff_2017}
---. 2017, Icarus, 293, 231, \dodoi{10.1016/j.icarus.2017.02.026}

\bibitem[{Griffith {et~al.}(2008)Griffith, McKay, \&
  Ferri}]{griffith_titans_2008}
Griffith, C.~A., McKay, C.~P., \& Ferri, F. 2008, The Astrophysical Journal
  Letters, 687, L41, \dodoi{10.1086/593117}

\bibitem[{Griffith {et~al.}(2003)Griffith, Owen, Geballe, Rayner, \&
  Rannou}]{griffith_evidence_2003}
Griffith, C.~A., Owen, T., Geballe, T.~R., Rayner, J., \& Rannou, P. 2003,
  Science, 300, 628, \dodoi{10.1126/science.1081897}

\bibitem[{Griffith {et~al.}(2019)Griffith, Penteado, Turner, Neish, Mitri,
  Montiel, Schoenfeld, \& Lopes}]{griffith_corridor_2019}
Griffith, C.~A., Penteado, P.~F., Turner, J.~D., {et~al.} 2019, Nature
  Astronomy, 3, 642, \dodoi{10.1038/s41550-019-0756-5}

\bibitem[{Guendelman \& Kaspi(2018)}]{guendelman_axisymmetric_2018}
Guendelman, I., \& Kaspi, Y. 2018, Geophysical Research Letters, 45, 13,213,
  \dodoi{https://doi.org/10.1029/2018GL080752}

\bibitem[{Gurwell(2004)}]{gurwell_submillimeter_2004}
Gurwell, M.~A. 2004, The Astrophysical Journal Letters, 616, L7,
  \dodoi{10.1086/423954}

\bibitem[{Hanley {et~al.}(2020)Hanley, Wing, Engle, Grundy, Lindberg, Tan, \&
  Tegler}]{hanley_effects_2020}
Hanley, J., Wing, B., Engle, A., {et~al.} 2020, Bulletin of the AAS, 52.
\newblock \url{https://baas.aas.org/pub/2020n6i408p06/release/1}

\bibitem[{Hartwig {et~al.}(2018)Hartwig, Meyerhofer, Lorenz, \&
  Lemmon}]{hartwig_analytical_2018}
Hartwig, J., Meyerhofer, P., Lorenz, R., \& Lemmon, E. 2018, Icarus, 299, 175,
  \dodoi{10.1016/j.icarus.2017.08.003}

\bibitem[{Hayes(2016)}]{hayes_lakes_2016}
Hayes, A.~G. 2016, Annual Review of Earth and Planetary Sciences, 44, 57,
  \dodoi{10.1146/annurev-earth-060115-012247}

\bibitem[{Hayes {et~al.}(2018)Hayes, Lorenz, \&
  Lunine}]{hayes_post-cassini_2018}
Hayes, A.~G., Lorenz, R.~D., \& Lunine, J.~I. 2018, Nature Geoscience, 11, 306,
  \dodoi{10.1038/s41561-018-0103-y}

\bibitem[{Hayes {et~al.}(2011)Hayes, Aharonson, Lunine, Kirk, Zebker, Wye,
  Lorenz, Turtle, Paillou, Mitri, Wall, Stofan, Mitchell, \&
  Elachi}]{hayes_transient_2011}
Hayes, A.~G., Aharonson, O., Lunine, J.~I., {et~al.} 2011, Icarus, 211, 655,
  \dodoi{10.1016/j.icarus.2010.08.017}

\bibitem[{Hayes {et~al.}(2017)Hayes, Birch, Dietrich, Howard, Kirk, Poggiali,
  Mastrogiuseppe, Michaelides, Corlies, Moore, Malaska, Mitchell, Lorenz, \&
  Wood}]{hayes_topographic_2017}
Hayes, A.~G., Birch, S. P.~D., Dietrich, W.~E., {et~al.} 2017, Geophysical
  Research Letters, 44, 11,745, \dodoi{https://doi.org/10.1002/2017GL075468}

\bibitem[{He {et~al.}(2017)He, H\"{o}rst, Riemer, Sebree, Pauley, \&
  Vuitton}]{he_carbon_2017}
He, C., H\"{o}rst, S.~M., Riemer, S., {et~al.} 2017, The Astrophysical Journal,
  841, L31, \dodoi{10.3847/2041-8213/aa74cc}

\bibitem[{Hedgepeth {et~al.}(2020)Hedgepeth, Neish, Turtle, Stiles, Kirk, \&
  Lorenz}]{hedgepeth_titans_2020}
Hedgepeth, J.~E., Neish, C.~D., Turtle, E.~P., {et~al.} 2020, Icarus, 344,
  113664, \dodoi{10.1016/j.icarus.2020.113664}

\bibitem[{Hemingway {et~al.}(2013)Hemingway, Nimmo, Zebker, \&
  Iess}]{hemingway_rigid_2013}
Hemingway, D., Nimmo, F., Zebker, H., \& Iess, L. 2013, Nature, 500, 550,
  \dodoi{10.1038/nature12400}

\bibitem[{Hickson {et~al.}(2014)Hickson, Loison, Cavalié, Hébrard, \&
  Dobrijevic}]{hickson_evolution_2014}
Hickson, K.~M., Loison, J.~C., Cavalié, T., Hébrard, E., \& Dobrijevic, M.
  2014, Astronomy \& Astrophysics, 572, A58,
  \dodoi{10.1051/0004-6361/201424703}

\bibitem[{Hodyss {et~al.}(2013)Hodyss, Choukroun, Sotin, \&
  Beauchamp}]{hodyss_solubility_2013}
Hodyss, R., Choukroun, M., Sotin, C., \& Beauchamp, P. 2013, Geophysical
  Research Letters, 40, 2935, \dodoi{https://doi.org/10.1002/grl.50630}

\bibitem[{Hoehler(2007)}]{hoehler_energy_2007}
Hoehler, T.~M. 2007, Astrobiology, 7, 824, \dodoi{10.1089/ast.2006.0095}

\bibitem[{Hofgartner {et~al.}(2014)Hofgartner, Hayes, Lunine, Zebker, Stiles,
  Sotin, Barnes, Turtle, Baines, Brown, Buratti, Clark, Encrenaz, Kirk,
  Le~Gall, Lopes, Lorenz, Malaska, Mitchell, Nicholson, Paillou, Radebaugh,
  Wall, \& Wood}]{hofgartner_transient_2014}
Hofgartner, J.~D., Hayes, A.~G., Lunine, J.~I., {et~al.} 2014, Nature
  Geoscience, 7, 493, \dodoi{10.1038/ngeo2190}

\bibitem[{Hofgartner {et~al.}(2016)Hofgartner, Hayes, Lunine, Zebker, Lorenz,
  Malaska, Mastrogiuseppe, Notarnicola, \& Soderblom}]{hofgartner_titans_2016}
---. 2016, Icarus, 271, 338, \dodoi{10.1016/j.icarus.2016.02.022}

\bibitem[{H\"{o}rst(2017)}]{horst_titans_2017}
H\"{o}rst, S.~M. 2017, Journal of Geophysical Research: Planets, 122, 432,
  \dodoi{https://doi.org/10.1002/2016JE005240}

\bibitem[{H\"{o}rst {et~al.}(2018)H\"{o}rst, He, Lewis, Kempton, Marley,
  Morley, Moses, Valenti, \& Vuitton}]{horst_haze_2018}
H\"{o}rst, S.~M., He, C., Lewis, N.~K., {et~al.} 2018, Nature Astronomy, 2,
  303, \dodoi{10.1038/s41550-018-0397-0}

\bibitem[{Iess {et~al.}(2012)Iess, Jacobson, Ducci, Stevenson, Lunine,
  Armstrong, Asmar, Racioppa, Rappaport, \& Tortora}]{iess_tides_2012}
Iess, L., Jacobson, R.~A., Ducci, M., {et~al.} 2012, Science, 337, 457,
  \dodoi{10.1126/science.1219631}

\bibitem[{Jennings {et~al.}(2009)Jennings, Flasar, Kunde, Samuelson, Pearl,
  Nixon, Carlson, Mamoutkine, Brasunas, Guandique, Achterberg, Bjoraker,
  Romani, Segura, Albright, Elliott, Tingley, Calcutt, Coustenis, \&
  Courtin}]{jennings_titans_2009}
Jennings, D.~E., Flasar, F.~M., Kunde, V.~G., {et~al.} 2009, The Astrophysical
  Journal Letters, 691, L103, \dodoi{10.1088/0004-637X/691/2/L103}

\bibitem[{Jennings {et~al.}(2011)Jennings, Cottini, Nixon, Flasar, Kunde,
  Samuelson, Romani, Hesman, Carlson, Gorius, Coustenis, \&
  Tokano}]{jennings_seasonal_2011}
Jennings, D.~E., Cottini, V., Nixon, C.~A., {et~al.} 2011, The Astrophysical
  Journal, 737, L15, \dodoi{10.1088/2041-8205/737/1/L15}

\bibitem[{Jennings {et~al.}(2012{\natexlab{a}})Jennings, Anderson, Samuelson,
  Flasar, Nixon, Bjoraker, Romani, Achterberg, Cottini, Hesman, Kunde, Carlson,
  Kok, Coustenis, Vinatier, Bampasidis, Teanby, \&
  Calcutt}]{jennings_first_2012}
Jennings, D.~E., Anderson, C.~M., Samuelson, R.~E., {et~al.}
  2012{\natexlab{a}}, The Astrophysical Journal, 761, L15,
  \dodoi{10.1088/2041-8205/761/1/L15}

\bibitem[{Jennings {et~al.}(2012{\natexlab{b}})Jennings, Anderson, Samuelson,
  Flasar, Nixon, Kunde, Achterberg, Cottini, Kok, Coustenis, Vinatier, \&
  Calcutt}]{jennings_seasonal_2012}
---. 2012{\natexlab{b}}, The Astrophysical Journal, 754, L3,
  \dodoi{10.1088/2041-8205/754/1/L3}

\bibitem[{Jennings {et~al.}(2015)Jennings, Achterberg, Cottini, Anderson,
  Flasar, Nixon, Bjoraker, Kunde, Carlson, Guandique, Kaelberer, Tingley,
  Albright, Segura, Kok, Coustenis, Vinatier, Bampasidis, Teanby, \&
  Calcutt}]{jennings_evolution_2015}
Jennings, D.~E., Achterberg, R.~K., Cottini, V., {et~al.} 2015, The
  Astrophysical Journal, 804, L34, \dodoi{10.1088/2041-8205/804/2/L34}

\bibitem[{Jennings {et~al.}(2016)Jennings, Cottini, Nixon, Achterberg, Flasar,
  Kunde, Romani, Samuelson, Mamoutkine, Gorius, Coustenis, \&
  Tokano}]{jennings_surface_2016}
Jennings, D.~E., Cottini, V., Nixon, C.~A., {et~al.} 2016, The Astrophysical
  Journal, 816, L17, \dodoi{10.3847/2041-8205/816/1/L17}

\bibitem[{Jennings {et~al.}(2019)Jennings, Tokano, Cottini, Nixon, Achterberg,
  Flasar, Kunde, Romani, Samuelson, Segura, Gorius, Guandique, Kaelberer, \&
  Coustenis}]{jennings_titan_2019}
Jennings, D.~E., Tokano, T., Cottini, V., {et~al.} 2019, The Astrophysical
  Journal, 877, L8, \dodoi{10.3847/2041-8213/ab1f91}

\bibitem[{Journaux {et~al.}(2013)Journaux, Daniel, Caracas, Montagnac, \&
  Cardon}]{journaux_influence_2013}
Journaux, B., Daniel, I., Caracas, R., Montagnac, G., \& Cardon, H. 2013,
  Icarus, 226, 355, \dodoi{10.1016/j.icarus.2013.05.039}

\bibitem[{Journaux {et~al.}(2017)Journaux, Daniel, Petitgirard, Cardon,
  Perrillat, Caracas, \& Mezouar}]{journaux_salt_2017}
Journaux, B., Daniel, I., Petitgirard, S., {et~al.} 2017, Earth and Planetary
  Science Letters, 463, 36, \dodoi{10.1016/j.epsl.2017.01.017}

\bibitem[{Journaux {et~al.}(2020{\natexlab{a}})Journaux, Brown, Pakhomova,
  Collings, Petitgirard, Espinoza, Ballaran, Vance, Ott, Cova, Garbarino, \&
  Hanfland}]{journaux_holistic_2020}
Journaux, B., Brown, J.~M., Pakhomova, A., {et~al.} 2020{\natexlab{a}}, Journal
  of Geophysical Research: Planets, 125, e2019JE006176,
  \dodoi{https://doi.org/10.1029/2019JE006176}

\bibitem[{Journaux {et~al.}(2020{\natexlab{b}})Journaux, Kalousov\'{a}, Sotin,
  Tobie, Vance, Saur, Bollengier, Noack, Rückriemen-Bez, Van~Hoolst,
  Soderlund, \& Brown}]{journaux_large_2020}
Journaux, B., Kalousov\'{a}, K., Sotin, C., {et~al.} 2020{\natexlab{b}}, Space
  Science Reviews, 216, \dodoi{10.1007/s11214-019-0633-7}

\bibitem[{Kalousov\'{a} \& Sotin(2018)}]{kalousova_melting_2018}
Kalousov\'{a}, K., \& Sotin, C. 2018, Geophysical Research Letters, 45, 8096,
  \dodoi{https://doi.org/10.1029/2018GL078889}

\bibitem[{Kalousov\'{a} {et~al.}(2018)Kalousov\'{a}, Sotin, Choblet, Tobie, \&
  Grasset}]{kalousova_two-phase_2018}
Kalousov\'{a}, K., Sotin, C., Choblet, G., Tobie, G., \& Grasset, O. 2018,
  Icarus, 299, 133, \dodoi{10.1016/j.icarus.2017.07.018}

\bibitem[{Karkoschka \& Schr\"{o}der(2016)}]{karkoschka_disr_2016}
Karkoschka, E., \& Schr\"{o}der, S.~E. 2016, Icarus, 270, 307,
  \dodoi{10.1016/j.icarus.2015.08.006}

\bibitem[{Kite {et~al.}(2020)Kite, Mischna, Gao, Yung, \&
  Turbet}]{kite_methane_2020}
Kite, E.~S., Mischna, M.~A., Gao, P., Yung, Y.~L., \& Turbet, M. 2020,
  Planetary and Space Science, 181, 104820, \dodoi{10.1016/j.pss.2019.104820}

\bibitem[{K\"{o}hn {et~al.}(2019)K\"{o}hn, Dujko, Chanrion, \&
  Neubert}]{kohn_streamer_2019}
K\"{o}hn, C., Dujko, S., Chanrion, O., \& Neubert, T. 2019, Icarus, 333, 294,
  \dodoi{10.1016/j.icarus.2019.05.036}

\bibitem[{Krasnopolsky(2009)}]{krasnopolsky_photochemical_2009}
Krasnopolsky, V.~A. 2009, Icarus, 201, 226,
  \dodoi{10.1016/j.icarus.2008.12.038}

\bibitem[{Krasnopolsky(2014)}]{krasnopolsky_chemical_2014}
---. 2014, Icarus, 236, 83, \dodoi{10.1016/j.icarus.2014.03.041}

\bibitem[{Lai {et~al.}(2017)Lai, Cordiner, Nixon, Achterberg, Molter, Teanby,
  Palmer, Charnley, Lindberg, Kisiel, Mumma, \& Irwin}]{lai_mapping_2017}
Lai, J. C.-Y., Cordiner, M.~A., Nixon, C.~A., {et~al.} 2017, The Astronomical
  Journal, 154, 206, \dodoi{10.3847/1538-3881/aa8eef}

\bibitem[{Langhans {et~al.}(2013)Langhans, Lunine, \&
  Mitri}]{langhans_titans_2013}
Langhans, M., Lunine, J.~I., \& Mitri, G. 2013, Icarus, 223, 796,
  \dodoi{10.1016/j.icarus.2013.01.016}

\bibitem[{Lara {et~al.}(2014)Lara, Lellouch, Gonz\'{a}lez, Moreno, \&
  Rengel}]{lara_time-dependent_2014}
Lara, L.~M., Lellouch, E., Gonz\'{a}lez, M., Moreno, R., \& Rengel, M. 2014,
  Astronomy \& Astrophysics, 566, A143, \dodoi{10.1051/0004-6361/201323085}

\bibitem[{Larson {et~al.}(2014)Larson, Toon, \&
  Friedson}]{larson_simulating_2014}
Larson, E. J.~L., Toon, O.~B., \& Friedson, A.~J. 2014, Icarus, 243, 400,
  \dodoi{10.1016/j.icarus.2014.09.003}

\bibitem[{Le~Gall {et~al.}(2014)Le~Gall, Janssen, Kirk, \&
  Lorenz}]{le_gall_modeling_2014}
Le~Gall, A., Janssen, M.~A., Kirk, R.~L., \& Lorenz, R.~D. 2014, Icarus, 230,
  198, \dodoi{10.1016/j.icarus.2013.06.009}

\bibitem[{Le~Gall {et~al.}(2010)Le~Gall, Janssen, Paillou, Lorenz, \&
  Wall}]{le_gall_radar-bright_2010}
Le~Gall, A., Janssen, M.~A., Paillou, P., Lorenz, R.~D., \& Wall, S.~D. 2010,
  Icarus, 207, 948, \dodoi{10.1016/j.icarus.2009.12.027}

\bibitem[{Le~Gall {et~al.}(2011)Le~Gall, Janssen, Wye, Hayes, Radebaugh,
  Savage, Zebker, Lorenz, Lunine, Kirk, Lopes, Wall, Callahan, Stofan, \&
  Farr}]{le_gall_cassini_2011}
Le~Gall, A., Janssen, M.~A., Wye, L.~C., {et~al.} 2011, Icarus, 213, 608,
  \dodoi{10.1016/j.icarus.2011.03.026}

\bibitem[{Le~Mouélic {et~al.}(2018)Le~Mouélic, Rodriguez, Robidel, Rousseau,
  Seignovert, Sotin, Barnes, Brown, Baines, Buratti, Clark, Nicholson, Rannou,
  \& Cornet}]{le_mouelic_mapping_2018}
Le~Mouélic, S., Rodriguez, S., Robidel, R., {et~al.} 2018, Icarus, 311, 371,
  \dodoi{10.1016/j.icarus.2018.04.028}

\bibitem[{Lefevre {et~al.}(2014)Lefevre, Tobie, Choblet, \&
  Čadek}]{lefevre_structure_2014}
Lefevre, A., Tobie, G., Choblet, G., \& Čadek, O. 2014, Icarus, 237, 16,
  \dodoi{10.1016/j.icarus.2014.04.006}

\bibitem[{Leitner \& Lunine(2019)}]{leitner_modeling_2019}
Leitner, M.~A., \& Lunine, J.~I. 2019, Icarus, 333, 61,
  \dodoi{10.1016/j.icarus.2019.05.008}

\bibitem[{Lellouch {et~al.}(2019)Lellouch, Gurwell, Moreno, Vinatier, Strobel,
  Moullet, Butler, Lara, Hidayat, \& Villard}]{lellouch_intense_2019}
Lellouch, E., Gurwell, M.~A., Moreno, R., {et~al.} 2019, Nature Astronomy, 3,
  614, \dodoi{10.1038/s41550-019-0749-4}

\bibitem[{Lemmon {et~al.}(2019)Lemmon, Lorenz, Smith, \&
  Caldwell}]{lemmon_large-scale_2019}
Lemmon, M.~T., Lorenz, R.~D., Smith, P.~H., \& Caldwell, J.~J. 2019, Icarus,
  331, 1, \dodoi{10.1016/j.icarus.2019.03.042}

\bibitem[{Levi \& Cohen(2019)}]{levi_equation_2019}
Levi, A., \& Cohen, R.~E. 2019, The Astrophysical Journal, 882, 71,
  \dodoi{10.3847/1538-4357/ab2f76}

\bibitem[{Lockwood {et~al.}(2008)Lockwood, Leary, Lorenz, Waite, Reh, Prince,
  \& Powell}]{lockwood2008titan}
Lockwood, M.~K., Leary, J., Lorenz, R., {et~al.} 2008, in AIAA/AAS
  Astrodynamics Specialist Conference and Exhibit, 7071

\bibitem[{Loison {et~al.}(2019)Loison, Dobrijevic, \&
  Hickson}]{loison_photochemical_2019}
Loison, J.~C., Dobrijevic, M., \& Hickson, K.~M. 2019, Icarus, 329, 55,
  \dodoi{10.1016/j.icarus.2019.03.024}

\bibitem[{Loison {et~al.}(2020)Loison, Wakelam, Gratier, \&
  Hickson}]{loison_gas-grain_2020}
Loison, J.-C., Wakelam, V., Gratier, P., \& Hickson, K.~M. 2020, Monthly
  Notices of the Royal Astronomical Society, 498, 4663,
  \dodoi{10.1093/mnras/staa2700}

\bibitem[{Loison {et~al.}(2015)Loison, Hébrard, Dobrijevic, Hickson, Caralp,
  Hue, Gronoff, Venot, \& Bénilan}]{loison_neutral_2015}
Loison, J.~C., Hébrard, E., Dobrijevic, M., {et~al.} 2015, Icarus, 247, 218,
  \dodoi{10.1016/j.icarus.2014.09.039}

\bibitem[{Lombardo {et~al.}(2019)Lombardo, Nixon, Greathouse, Bézard, Jolly,
  Vinatier, Teanby, Richter, Irwin, Coustenis, \&
  Flasar}]{lombardo_detection_2019}
Lombardo, N.~A., Nixon, C.~A., Greathouse, T.~K., {et~al.} 2019, The
  Astrophysical Journal, 881, L33, \dodoi{10.3847/2041-8213/ab3860}

\bibitem[{Lopes {et~al.}(2007)Lopes, Mitchell, Stofan, Lunine, Lorenz,
  Paganelli, Kirk, Wood, Wall, Robshaw, Fortes, Neish, Radebaugh, Reffet,
  Ostro, Elachi, Allison, Anderson, Boehmer, Boubin, Callahan, Encrenaz,
  Flamini, Francescetti, Gim, Hamilton, Hensley, Janssen, Johnson, Kelleher,
  Muhleman, Ori, Orosei, Picardi, Posa, Roth, Seu, Shaffer, Soderblom, Stiles,
  Vetrella, West, Wye, \& Zebker}]{lopes_cryovolcanic_2007}
Lopes, R. M.~C., Mitchell, K.~L., Stofan, E.~R., {et~al.} 2007, Icarus, 186,
  395, \dodoi{10.1016/j.icarus.2006.09.006}

\bibitem[{Lopes {et~al.}(2013)Lopes, Kirk, Mitchell, LeGall, Barnes, Hayes,
  Kargel, Wye, Radebaugh, Stofan, Janssen, Neish, Wall, Wood, Lunine, \&
  Malaska}]{lopes_cryovolcanism_2013}
Lopes, R. M.~C., Kirk, R.~L., Mitchell, K.~L., {et~al.} 2013, Journal of
  Geophysical Research: Planets, 118, 416,
  \dodoi{https://doi.org/10.1002/jgre.20062}

\bibitem[{Lopes {et~al.}(2016)Lopes, Malaska, Solomonidou, Le~Gall, Janssen,
  Neish, Turtle, Birch, Hayes, Radebaugh, Coustenis, Schoenfeld, Stiles, Kirk,
  Mitchell, Stofan, \& Lawrence}]{lopes_nature_2016}
Lopes, R. M.~C., Malaska, M.~J., Solomonidou, A., {et~al.} 2016, Icarus, 270,
  162, \dodoi{10.1016/j.icarus.2015.11.034}

\bibitem[{Lopes {et~al.}(2020)Lopes, Malaska, Schoenfeld, Solomonidou, Birch,
  Florence, Hayes, Williams, Radebaugh, Verlander, Turtle, Le~Gall, \&
  Wall}]{lopes_global_2020}
Lopes, R. M.~C., Malaska, M.~J., Schoenfeld, A.~M., {et~al.} 2020, Nature
  Astronomy, 4, 228, \dodoi{10.1038/s41550-019-0917-6}

\bibitem[{Lora {et~al.}(2018)Lora, Kataria, \& Gao}]{lora_atmospheric_2018}
Lora, J.~M., Kataria, T., \& Gao, P. 2018, The Astrophysical Journal, 853, 58,
  \dodoi{10.3847/1538-4357/aaa132}

\bibitem[{Lora {et~al.}(2015)Lora, Lunine, \& Russell}]{lora_gcm_2015}
Lora, J.~M., Lunine, J.~I., \& Russell, J.~L. 2015, Icarus, 250, 516,
  \dodoi{10.1016/j.icarus.2014.12.030}

\bibitem[{Lora {et~al.}(2014)Lora, Lunine, Russell, \&
  Hayes}]{lora_simulations_2014}
Lora, J.~M., Lunine, J.~I., Russell, J.~L., \& Hayes, A.~G. 2014, Icarus, 243,
  264, \dodoi{10.1016/j.icarus.2014.08.042}

\bibitem[{Lora \& Mitchell(2015)}]{lora_titans_2015}
Lora, J.~M., \& Mitchell, J.~L. 2015, Geophysical Research Letters, 42, 6213,
  \dodoi{https://doi.org/10.1002/2015GL064912}

\bibitem[{Lora {et~al.}(2019)Lora, Tokano, Vatant~d’Ollone, Lebonnois, \&
  Lorenz}]{lora_model_2019}
Lora, J.~M., Tokano, T., Vatant~d’Ollone, J., Lebonnois, S., \& Lorenz, R.~D.
  2019, Icarus, 333, 113, \dodoi{10.1016/j.icarus.2019.05.031}

\bibitem[{Lorenz(2008)}]{lorenz_review_2008}
Lorenz, R.~D. 2008, Journal of the British Interplanetary Society, 61, 2.
\newblock \url{http://adsabs.harvard.edu/abs/2008JBIS...61....2L}

\bibitem[{Lorenz {et~al.}(2008{\natexlab{a}})Lorenz, Leary, Lockwood, \&
  Waite}]{lorenz2008titan}
Lorenz, R.~D., Leary, J.~C., Lockwood, M.~K., \& Waite, J.~H.
  2008{\natexlab{a}}, in AIP Conference Proceedings, Vol. 969, American
  Institute of Physics, 380--387

\bibitem[{Lorenz {et~al.}(2016)Lorenz, Oleson, Colozza, Hartwig, Schmitz,
  Landis, Paul, \& Walsh}]{lorenz_submarine_2016}
Lorenz, R.~D., Oleson, S., Colozza, T., {et~al.} 2016, in EGU General Assembly
  Conference, Vol.~18, EPSC2016--5455.
\newblock \url{http://adsabs.harvard.edu/abs/2016EGUGA..18.5455L}

\bibitem[{Lorenz \& Radebaugh(2009)}]{lorenz_global_2009}
Lorenz, R.~D., \& Radebaugh, J. 2009, Geophysical Research Letters, 36,
  \dodoi{https://doi.org/10.1029/2008GL036850}

\bibitem[{Lorenz {et~al.}(2007)Lorenz, Wood, Lunine, Wall, Lopes, Mitchell,
  Paganelli, Anderson, Wye, Tsai, Zebker, \& Stofan}]{lorenz_titans_2007}
Lorenz, R.~D., Wood, C.~A., Lunine, J.~I., {et~al.} 2007, Geophysical Research
  Letters, 34, \dodoi{https://doi.org/10.1029/2006GL028971}

\bibitem[{Lorenz {et~al.}(2008{\natexlab{b}})Lorenz, Mitchell, Kirk, Hayes,
  Aharonson, Zebker, Paillou, Radebaugh, Lunine, Janssen, Wall, Lopes, Stiles,
  Ostro, Mitri, \& Stofan}]{lorenz_titans_2008}
Lorenz, R.~D., Mitchell, K.~L., Kirk, R.~L., {et~al.} 2008{\natexlab{b}},
  Geophysical Research Letters, 35,
  \dodoi{https://doi.org/10.1029/2007GL032118}

\bibitem[{Lorenz {et~al.}(2018{\natexlab{a}})Lorenz, Turtle, Barnes, Trainer,
  Adams, Hibbard, Sheldon, Zacny, Peplowski, Lawrence, \&
  {others}}]{lorenz_dragonfly_2018}
Lorenz, R.~D., Turtle, E.~P., Barnes, J.~W., {et~al.} 2018{\natexlab{a}}, Johns
  Hopkins APL Technical Digest, 34, 14

\bibitem[{Lorenz {et~al.}(2018{\natexlab{b}})Lorenz, Oleson, Colozza, Jones,
  Packard, Hartwig, Newman, Gyekenyesi, Schmitz, \&
  Walsh}]{lorenz_dropsonde_2018}
Lorenz, R.~D., Oleson, S.~R., Colozza, A.~J., {et~al.} 2018{\natexlab{b}},
  Advances in Space Research, 62, 912, \dodoi{10.1016/j.asr.2018.05.030}

\bibitem[{Lunine(2010)}]{lunine_titan_2010}
Lunine, J.~I. 2010, Faraday Discussions, 147, 405, \dodoi{10.1039/C004788K}

\bibitem[{Lunine \& Lorenz(2009)}]{lunine_rivers_2009}
Lunine, J.~I., \& Lorenz, R.~D. 2009, Annual Review of Earth and Planetary
  Sciences, 37, 299, \dodoi{10.1146/annurev.earth.031208.100142}

\bibitem[{Lunine \& Stevenson(1987)}]{lunine_clathrate_1987}
Lunine, J.~I., \& Stevenson, D.~J. 1987, Icarus, 70, 61,
  \dodoi{10.1016/0019-1035(87)90075-3}

\bibitem[{Lunine {et~al.}(1983)Lunine, Stevenson, \& Yung}]{lunine_ethane_1983}
Lunine, J.~I., Stevenson, D.~J., \& Yung, Y.~L. 1983, Science, 222, 1229,
  \dodoi{10.1126/science.222.4629.1229}

\bibitem[{Luspay-Kuti {et~al.}(2015)Luspay-Kuti, Mandt, Plessis, \&
  Greathouse}]{luspay-kuti_effects_2015}
Luspay-Kuti, A., Mandt, K.~E., Plessis, S., \& Greathouse, T.~K. 2015, The
  Astrophysical Journal, 801, L14, \dodoi{10.1088/2041-8205/801/1/L14}

\bibitem[{Lv {et~al.}(2017)Lv, Norman, \& Li}]{lv_oxygen-free_2017}
Lv, K.-P., Norman, L., \& Li, Y.-L. 2017, Astrobiology, 17, 1173,
  \dodoi{10.1089/ast.2016.1574}

\bibitem[{MacKenzie {et~al.}(2019{\natexlab{a}})MacKenzie, Lora, \&
  Lorenz}]{mackenzie_thermal_2019}
MacKenzie, S.~M., Lora, J.~M., \& Lorenz, R.~D. 2019{\natexlab{a}}, Journal of
  Geophysical Research (Planets), 124, 1728, \dodoi{10.1029/2019JE005930}

\bibitem[{MacKenzie {et~al.}(2014)MacKenzie, Barnes, Sotin, Soderblom,
  Le~Mouélic, Rodriguez, Baines, Buratti, Clark, Nicholson, \&
  McCord}]{mackenzie_evidence_2014}
MacKenzie, S.~M., Barnes, J.~W., Sotin, C., {et~al.} 2014, Icarus, 243, 191,
  \dodoi{10.1016/j.icarus.2014.08.022}

\bibitem[{MacKenzie {et~al.}(2019{\natexlab{b}})MacKenzie, Barnes, Hofgartner,
  Birch, Hedman, Lucas, Rodriguez, Turtle, \& Sotin}]{mackenzie_case_2019}
MacKenzie, S.~M., Barnes, J.~W., Hofgartner, J.~D., {et~al.}
  2019{\natexlab{b}}, Nature Astronomy, 3, 506,
  \dodoi{10.1038/s41550-018-0687-6}

\bibitem[{Malaska {et~al.}(2017)Malaska, Hodyss, Lunine, Hayes, Hofgartner,
  Hollyday, \& Lorenz}]{malaska_laboratory_2017}
Malaska, M.~J., Hodyss, R., Lunine, J.~I., {et~al.} 2017, Icarus, 289, 94,
  \dodoi{10.1016/j.icarus.2017.01.033}

\bibitem[{Malaska {et~al.}(2016{\natexlab{a}})Malaska, Lopes, Hayes, Radebaugh,
  Lorenz, \& Turtle}]{malaska_material_2016}
Malaska, M.~J., Lopes, R.~M., Hayes, A.~G., {et~al.} 2016{\natexlab{a}},
  Icarus, 270, 183, \dodoi{10.1016/j.icarus.2015.09.029}

\bibitem[{Malaska {et~al.}(2016{\natexlab{b}})Malaska, Lopes, Williams, Neish,
  Solomonidou, Soderblom, Schoenfeld, Birch, Hayes, Le~Gall, Janssen, Farr,
  Lorenz, Radebaugh, \& Turtle}]{malaska_geomorphological_2016}
Malaska, M.~J., Lopes, R. M.~C., Williams, D.~A., {et~al.} 2016{\natexlab{b}},
  Icarus, 270, 130, \dodoi{10.1016/j.icarus.2016.02.021}

\bibitem[{Mandt {et~al.}(2015)Mandt, Mousis, \&
  Chassefière}]{mandt_comparative_2015}
Mandt, K., Mousis, O., \& Chassefière, E. 2015, Icarus, 254, 259,
  \dodoi{10.1016/j.icarus.2015.03.025}

\bibitem[{Marounina {et~al.}(2018)Marounina, Grasset, Tobie, \&
  Carpy}]{marounina_role_2018}
Marounina, N., Grasset, O., Tobie, G., \& Carpy, S. 2018, Icarus, 310, 127,
  \dodoi{10.1016/j.icarus.2017.10.048}

\bibitem[{Marounina {et~al.}(2015)Marounina, Tobie, Carpy, Monteux, Charnay, \&
  Grasset}]{marounina_evolution_2015}
Marounina, N., Tobie, G., Carpy, S., {et~al.} 2015, Icarus, 257, 324,
  \dodoi{10.1016/j.icarus.2015.05.011}

\bibitem[{Marten {et~al.}(2002)Marten, Hidayat, Biraud, \&
  Moreno}]{marten_new_2002}
Marten, A., Hidayat, T., Biraud, Y., \& Moreno, R. 2002, Icarus, 158, 532,
  \dodoi{10.1006/icar.2002.6897}

\bibitem[{Mart\'{i}nez-Rodr\'{i}guez {et~al.}(2019)Mart\'{i}nez-Rodr\'{i}guez,
  Caballero, Cifuentes, Piro, \& Barnes}]{martinez-rodriguez_exomoons_2019}
Mart\'{i}nez-Rodr\'{i}guez, H., Caballero, J.~A., Cifuentes, C., Piro, A.~L.,
  \& Barnes, R. 2019, The Astrophysical Journal, 887, 261,
  \dodoi{10.3847/1538-4357/ab5640}

\bibitem[{Mastrogiuseppe {et~al.}(2016)Mastrogiuseppe, Hayes, Poggiali, Seu,
  Lunine, \& Hofgartner}]{mastrogiuseppe_radar_2016}
Mastrogiuseppe, M., Hayes, A., Poggiali, V., {et~al.} 2016, IEEE Transactions
  on Geoscience and Remote Sensing, 54, 5646, \dodoi{10.1109/TGRS.2016.2563426}

\bibitem[{Mastrogiuseppe {et~al.}(2014)Mastrogiuseppe, Poggiali, Hayes, Lorenz,
  Lunine, Picardi, Seu, Flamini, Mitri, Notarnicola, Paillou, \&
  Zebker}]{mastrogiuseppe_bathymetry_2014}
Mastrogiuseppe, M., Poggiali, V., Hayes, A., {et~al.} 2014, Geophysical
  Research Letters, 41, 1432, \dodoi{https://doi.org/10.1002/2013GL058618}

\bibitem[{Mastrogiuseppe {et~al.}(2018)Mastrogiuseppe, Hayes, Poggiali, Lunine,
  Lorenz, Seu, Le~Gall, Notarnicola, Mitchell, Malaska, \&
  Birch}]{mastrogiuseppe_bathymetry_2018}
Mastrogiuseppe, M., Hayes, A.~G., Poggiali, V., {et~al.} 2018, Icarus, 300,
  203, \dodoi{10.1016/j.icarus.2017.09.009}

\bibitem[{Maynard-Casely {et~al.}(2018)Maynard-Casely, Cable, Malaska, Vu,
  Choukroun, \& Hodyss}]{maynard-casely_prospects_2018}
Maynard-Casely, H.~E., Cable, M.~L., Malaska, M.~J., {et~al.} 2018, American
  Mineralogist, 103, 343, \dodoi{10.2138/am-2018-6259}

\bibitem[{McKay(2016)}]{mckay_titan_2016}
McKay, C.~P. 2016, Life, 6, 8, \dodoi{10.3390/life6010008}

\bibitem[{Meier {et~al.}(2000)Meier, Smith, Owen, \&
  Terrile}]{meier_surface_2000}
Meier, R., Smith, B.~A., Owen, T.~C., \& Terrile, R.~J. 2000, Icarus, 145, 462,
  \dodoi{10.1006/icar.2000.6360}

\bibitem[{Miguel(2019)}]{miguel_observability_2019}
Miguel, Y. 2019, Monthly Notices of the Royal Astronomical Society, 482, 2893,
  \dodoi{10.1093/mnras/sty2803}

\bibitem[{Miller {et~al.}(2019)Miller, Glein, \&
  Waite}]{miller_contributions_2019}
Miller, K.~E., Glein, C.~R., \& Waite, J.~H. 2019, The Astrophysical Journal,
  871, 59, \dodoi{10.3847/1538-4357/aaf561}

\bibitem[{Mitchell(2008)}]{mitchell_drying_2008}
Mitchell, J.~L. 2008, Journal of Geophysical Research: Planets, 113,
  \dodoi{https://doi.org/10.1029/2007JE003017}

\bibitem[{Mitchell \& Lora(2016)}]{mitchell_climate_2016}
Mitchell, J.~L., \& Lora, J.~M. 2016, Annual Review of Earth and Planetary
  Sciences, 44, 353, \dodoi{10.1146/annurev-earth-060115-012428}

\bibitem[{Mitchell {et~al.}(2006)Mitchell, Pierrehumbert, Frierson, \&
  Caballero}]{mitchell_dynamics_2006}
Mitchell, J.~L., Pierrehumbert, R.~T., Frierson, D. M.~W., \& Caballero, R.
  2006, Proceedings of the National Academy of Sciences, 103, 18421,
  \dodoi{10.1073/pnas.0605074103}

\bibitem[{Mitchell {et~al.}(2009)Mitchell, Pierrehumbert, Frierson, \&
  Caballero}]{mitchell_impact_2009}
---. 2009, Icarus, 203, 250, \dodoi{10.1016/j.icarus.2009.03.043}

\bibitem[{Mitchell {et~al.}(2014)Mitchell, Vallis, \&
  Potter}]{mitchell_effects_2014}
Mitchell, J.~L., Vallis, G.~K., \& Potter, S.~F. 2014, The Astrophysical
  Journal, 787, 23, \dodoi{10.1088/0004-637X/787/1/23}

\bibitem[{Miteva {et~al.}(2009)Miteva, Teacher, Sowers, \&
  Brenchley}]{miteva_comparison_2009}
Miteva, V., Teacher, C., Sowers, T., \& Brenchley, J. 2009, Environmental
  Microbiology, 11, 640,
  \dodoi{https://doi.org/10.1111/j.1462-2920.2008.01835.x}

\bibitem[{Mitri {et~al.}(2014)Mitri, Meriggiola, Hayes, Lefevre, Tobie, Genova,
  Lunine, \& Zebker}]{mitri_shape_2014}
Mitri, G., Meriggiola, R., Hayes, A., {et~al.} 2014, Icarus, 236, 169,
  \dodoi{10.1016/j.icarus.2014.03.018}

\bibitem[{Mitri {et~al.}(2008)Mitri, Showman, Lunine, \&
  Lopes}]{mitri_resurfacing_2008}
Mitri, G., Showman, A.~P., Lunine, J.~I., \& Lopes, R. M.~C. 2008, Icarus, 196,
  216, \dodoi{10.1016/j.icarus.2008.02.024}

\bibitem[{Molter {et~al.}(2016)Molter, Nixon, Cordiner, Serigano, Irwin,
  Teanby, Charnley, \& Lindberg}]{molter_alma_2016}
Molter, E.~M., Nixon, C.~A., Cordiner, M.~A., {et~al.} 2016, The Astronomical
  Journal, 152, 42, \dodoi{10.3847/0004-6256/152/2/42}

\bibitem[{Moore \& Pappalardo(2011)}]{moore_titan_2011}
Moore, J.~M., \& Pappalardo, R.~T. 2011, Icarus, 212, 790,
  \dodoi{10.1016/j.icarus.2011.01.019}

\bibitem[{Morley {et~al.}(2017)Morley, Kreidberg, Rustamkulov, Robinson, \&
  Fortney}]{morley2017observing}
Morley, C.~V., Kreidberg, L., Rustamkulov, Z., Robinson, T., \& Fortney, J.~J.
  2017, The Astrophysical Journal, 850, 121

\bibitem[{Mukundan \& Bhardwaj(2018)}]{mukundan_model_2018}
Mukundan, V., \& Bhardwaj, A. 2018, The Astrophysical Journal, 856, 168,
  \dodoi{10.3847/1538-4357/aab1f5}

\bibitem[{Mu\{{\textbackslash}{\textasciitilde}n\}oz
  {et~al.}(2017)Mu\{{\textbackslash}{\textasciitilde}n\}oz, Lavvas, \&
  West}]{munoz_titan_2017}
Mu\{{\textbackslash}{\textasciitilde}n\}oz, A.~G., Lavvas, P., \& West, R.~A.
  2017, Nature Astronomy, 1, 1, \dodoi{10.1038/s41550-017-0114}

\bibitem[{Méndez~Harper {et~al.}(2017)Méndez~Harper, McDonald, Dufek,
  Malaska, Burr, Hayes, McAdams, \& Wray}]{mendez_harper_electrification_2017}
Méndez~Harper, J.~S., McDonald, G.~D., Dufek, J., {et~al.} 2017, Nature
  Geoscience, 10, 260, \dodoi{10.1038/ngeo2921}

\bibitem[{{National Research
  Council}(2011)}]{national_research_council_vision_2011}
{National Research Council}. 2011, Vision and {Voyages} for {Planetary}
  {Science} in the {Decade} 2013-2022 (Washington, DC: The National Academies
  Press), \dodoi{10.17226/13117}

\bibitem[{Neish {et~al.}(2008)Neish, Somogyi, Imanaka, Lunine, \&
  Smith}]{neish_rate_2008}
Neish, C., Somogyi, A., Imanaka, H., Lunine, J., \& Smith, M. 2008,
  Astrobiology, 8, 273, \dodoi{10.1089/ast.2007.0193}

\bibitem[{Neish \& Lorenz(2012)}]{neish_titans_2012}
Neish, C.~D., \& Lorenz, R.~D. 2012, Planetary and Space Science, 60, 26,
  \dodoi{10.1016/j.pss.2011.02.016}

\bibitem[{Neish \& Lorenz(2014)}]{neish_elevation_2014}
---. 2014, Icarus, 228, 27, \dodoi{10.1016/j.icarus.2013.09.024}

\bibitem[{Neish {et~al.}(2006)Neish, Lorenz, \& O'Brien}]{neish_potential_2006}
Neish, C.~D., Lorenz, R.~D., \& O'Brien, D.~P. 2006, International Journal of
  Astrobiology, 5, 57, \dodoi{10.1017/S1473550406002898}

\bibitem[{Neish {et~al.}(2016)Neish, Molaro, Lora, Howard, Kirk, Schenk, Bray,
  \& Lorenz}]{neish_fluvial_2016}
Neish, C.~D., Molaro, J.~L., Lora, J.~M., {et~al.} 2016, Icarus, 270, 114,
  \dodoi{10.1016/j.icarus.2015.07.022}

\bibitem[{Neish {et~al.}(2009)Neish, Somogyi, Lunine, \&
  Smith}]{neish_low_2009}
Neish, C.~D., Somogyi, A., Lunine, J.~I., \& Smith, M.~A. 2009, Icarus, 201,
  412, \dodoi{10.1016/j.icarus.2009.01.003}

\bibitem[{Neish {et~al.}(2010)Neish, Somogyi, \& Smith}]{neish_titans_2010}
Neish, C.~D., Somogyi, A., \& Smith, M.~A. 2010, Astrobiology, 10, 337,
  \dodoi{10.1089/ast.2009.0402}

\bibitem[{Neish {et~al.}(2013)Neish, Kirk, Lorenz, Bray, Schenk, Stiles,
  Turtle, Mitchell, \& Hayes}]{neish_crater_2013}
Neish, C.~D., Kirk, R.~L., Lorenz, R.~D., {et~al.} 2013, Icarus, 223, 82,
  \dodoi{10.1016/j.icarus.2012.11.030}

\bibitem[{Neish {et~al.}(2018)Neish, Lorenz, Turtle, Barnes, Trainer, Stiles,
  Kirk, Hibbitts, \& Malaska}]{neish_strategies_2018}
Neish, C.~D., Lorenz, R.~D., Turtle, E.~P., {et~al.} 2018, Astrobiology, 18,
  571, \dodoi{10.1089/ast.2017.1758}

\bibitem[{Neveu \& Rhoden(2019)}]{neveu_evolution_2019}
Neveu, M., \& Rhoden, A.~R. 2019, Nature Astronomy, 3, 543,
  \dodoi{10.1038/s41550-019-0726-y}

\bibitem[{Newman {et~al.}(2016)Newman, Richardson, Lian, \&
  Lee}]{newman_simulating_2016}
Newman, C.~E., Richardson, M.~I., Lian, Y., \& Lee, C. 2016, Icarus, 267, 106,
  \dodoi{10.1016/j.icarus.2015.11.028}

\bibitem[{Niemann {et~al.}(2005)Niemann, Atreya, Bauer, Carignan, Demick,
  Frost, Gautier, Haberman, Harpold, Hunten, Israel, Lunine, Kasprzak, Owen,
  Paulkovich, Raulin, Raaen, \& Way}]{niemann_abundances_2005}
Niemann, H.~B., Atreya, S.~K., Bauer, S.~J., {et~al.} 2005, Nature, 438, 779,
  \dodoi{10.1038/nature04122}

\bibitem[{Nimmo \& Pappalardo(2016)}]{nimmo_ocean_2016}
Nimmo, F., \& Pappalardo, R.~T. 2016, Journal of Geophysical Research: Planets,
  121, 1378, \dodoi{https://doi.org/10.1002/2016JE005081}

\bibitem[{Nixon {et~al.}(2012)Nixon, Temelso, Vinatier, Teanby, Bézard,
  Achterberg, Mandt, Sherrill, Irwin, Jennings, Romani, Coustenis, \&
  Flasar}]{nixon_isotopic_2012}
Nixon, C.~A., Temelso, B., Vinatier, S., {et~al.} 2012, The Astrophysical
  Journal, 749, 159, \dodoi{10.1088/0004-637X/749/2/159}

\bibitem[{Nixon {et~al.}(2018)Nixon, Lorenz, Achterberg, Buch, Coll, Clark,
  Courtin, Hayes, Iess, Johnson, Lopes, Mastrogiuseppe, Mandt, Mitchell,
  Raulin, Rymer, Todd~Smith, Solomonidou, Sotin, Strobel, Turtle, Vuitton,
  West, \& Yelle}]{nixon_titans_2018}
Nixon, C.~A., Lorenz, R.~D., Achterberg, R.~K., {et~al.} 2018, Planetary and
  Space Science, 155, 50, \dodoi{10.1016/j.pss.2018.02.009}

\bibitem[{Nixon {et~al.}(2020)Nixon, Thelen, Cordiner, Kisiel, Charnley,
  Molter, Serigano, Irwin, Teanby, \& Kuan}]{nixon_detection_2020}
Nixon, C.~A., Thelen, A.~E., Cordiner, M.~A., {et~al.} 2020, The Astronomical
  Journal, 160, 205, \dodoi{10.3847/1538-3881/abb679}

\bibitem[{Nna-Mvondo {et~al.}(2019)Nna-Mvondo, Anderson, \&
  Samuelson}]{nna-mvondo_detailed_2019}
Nna-Mvondo, D., Anderson, C.~M., \& Samuelson, R.~E. 2019, Icarus, 333, 183,
  \dodoi{10.1016/j.icarus.2019.05.003}

\bibitem[{Noguchi \& Okuchi(2020)}]{noguchi_rheological_2020}
Noguchi, N., \& Okuchi, T. 2020, Icarus, 335, 113401,
  \dodoi{10.1016/j.icarus.2019.113401}

\bibitem[{O'Brien {et~al.}(2005)O'Brien, Lorenz, \&
  Lunine}]{obrien_numerical_2005}
O'Brien, D.~P., Lorenz, R.~D., \& Lunine, J.~I. 2005, Icarus, 173, 243,
  \dodoi{10.1016/j.icarus.2004.08.001}

\bibitem[{Orgel(1998)}]{orgel_origin_1998}
Orgel, L.~E. 1998, Origins of life and evolution of the biosphere, 28, 91,
  \dodoi{10.1023/A:1006561308498}

\bibitem[{O’Rourke \& Stevenson(2014)}]{orourke_stability_2014}
O’Rourke, J.~G., \& Stevenson, D.~J. 2014, Icarus, 227, 67,
  \dodoi{10.1016/j.icarus.2013.09.010}

\bibitem[{Paillou {et~al.}(2016)Paillou, Seignovert, Radebaugh, \&
  Wall}]{paillou_radar_2016}
Paillou, P., Seignovert, B., Radebaugh, J., \& Wall, S. 2016, Icarus, 270, 211,
  \dodoi{10.1016/j.icarus.2015.07.038}

\bibitem[{Palmer {et~al.}(2017)Palmer, Cordiner, Nixon, Charnley, Teanby,
  Kisiel, Irwin, \& Mumma}]{palmer_alma_2017}
Palmer, M.~Y., Cordiner, M.~A., Nixon, C.~A., {et~al.} 2017, Science Advances,
  3, e1700022, \dodoi{10.1126/sciadv.1700022}

\bibitem[{Poggiali {et~al.}(2016)Poggiali, Mastrogiuseppe, Hayes, Seu, Birch,
  Lorenz, Grima, \& Hofgartner}]{poggiali_liquid-filled_2016}
Poggiali, V., Mastrogiuseppe, M., Hayes, A.~G., {et~al.} 2016, Geophysical
  Research Letters, 43, 7887, \dodoi{https://doi.org/10.1002/2016GL069679}

\bibitem[{Price(2007)}]{price_microbial_2007}
Price, P.~B. 2007, FEMS Microbiology Ecology, 59, 217,
  \dodoi{https://doi.org/10.1111/j.1574-6941.2006.00234.x}

\bibitem[{Radebaugh {et~al.}(2010)Radebaugh, Lorenz, Farr, Paillou, Savage, \&
  Spencer}]{radebaugh_linear_2010}
Radebaugh, J., Lorenz, R., Farr, T., {et~al.} 2010, Geomorphology, 121, 122,
  \dodoi{10.1016/j.geomorph.2009.02.022}

\bibitem[{Radebaugh {et~al.}(2008)Radebaugh, Lorenz, Lunine, Wall, Boubin,
  Reffet, Kirk, Lopes, Stofan, Soderblom, Allison, Janssen, Paillou, Callahan,
  Spencer, \& {the Cassini Radar Team}}]{radebaugh_dunes_2008}
Radebaugh, J., Lorenz, R.~D., Lunine, J.~I., {et~al.} 2008, Icarus, 194, 690,
  \dodoi{10.1016/j.icarus.2007.10.015}

\bibitem[{Radebaugh {et~al.}(2018)Radebaugh, Ventra, Lorenz, Farr, Kirk, Hayes,
  Malaska, Birch, Liu, Lunine, Barnes, Le~Gall, Lopes, Stofan, Wall, \&
  Paillou}]{radebaugh_alluvial_2018}
Radebaugh, J., Ventra, D., Lorenz, R.~D., {et~al.} 2018, in Geology and
  {Geomorphology} of {Alluvial} and {Fluvial} {Fans}: {Terrestrial} and
  {Planetary} {Perspectives}, ed. D.~Ventra \& L.~E. Clarke, Vol. 440
  (Geological Society of London), 0, \dodoi{10.1144/SP440.6}

\bibitem[{Rafkin \& Soto(2020)}]{rafkin_air-sea_2020}
Rafkin, S. C.~R., \& Soto, A. 2020, Icarus, 351, 113903,
  \dodoi{10.1016/j.icarus.2020.113903}

\bibitem[{Rahm {et~al.}(2016)Rahm, Lunine, Usher, \&
  Shalloway}]{rahm_polymorphism_2016}
Rahm, M., Lunine, J.~I., Usher, D.~A., \& Shalloway, D. 2016, Proceedings of
  the National Academy of Science, 113, 8121, \dodoi{10.1073/pnas.1606634113}

\bibitem[{Rannou {et~al.}(2004)Rannou, Hourdin, McKay, \&
  Luz}]{rannou_coupled_2004}
Rannou, P., Hourdin, F., McKay, C.~P., \& Luz, D. 2004, Icarus, 170, 443,
  \dodoi{10.1016/j.icarus.2004.03.007}

\bibitem[{Read \& Lebonnois(2018)}]{read_superrotation_2018}
Read, P.~L., \& Lebonnois, S. 2018, Annual Review of Earth and Planetary
  Sciences, 46, 175, \dodoi{10.1146/annurev-earth-082517-010137}

\bibitem[{Reh(2009)}]{reh2009titan}
Reh, K.~R. 2009, in 2009 IEEE Aerospace conference, IEEE, 1--8

\bibitem[{Robinson {et~al.}(2014)Robinson, Maltagliati, Marley, \&
  Fortney}]{robinson_titan_2014}
Robinson, T.~D., Maltagliati, L., Marley, M.~S., \& Fortney, J.~J. 2014,
  Proceedings of the National Academy of Sciences, 111, 9042,
  \dodoi{10.1073/pnas.1403473111}

\bibitem[{Rodriguez(in review)}]{rodriguez_poseidon}
Rodriguez, S. in review, Experimental Astronomy

\bibitem[{Rodriguez {et~al.}(2014)Rodriguez, Garcia, Lucas, Appéré, Le~Gall,
  Reffet, Le~Corre, Le~Mouélic, Cornet, Courrech~du Pont, Narteau, Bourgeois,
  Radebaugh, Arnold, Barnes, Stephan, Jaumann, Sotin, Brown, Lorenz, \&
  Turtle}]{rodriguez_global_2014}
Rodriguez, S., Garcia, A., Lucas, A., {et~al.} 2014, Icarus, 230, 168,
  \dodoi{10.1016/j.icarus.2013.11.017}

\bibitem[{Rodriguez {et~al.}(2018)Rodriguez, Le~Mouélic, Barnes, Kok, Rafkin,
  Lorenz, Charnay, Radebaugh, Narteau, Cornet, Bourgeois, Lucas, Rannou,
  Griffith, Coustenis, Appéré, Hirtzig, Sotin, Soderblom, Brown, Bow, Vixie,
  Maltagliati, Courrech~du Pont, Jaumann, Stephan, Baines, Buratti, Clark, \&
  Nicholson}]{rodriguez_observational_2018}
Rodriguez, S., Le~Mouélic, S., Barnes, J.~W., {et~al.} 2018, Nature
  Geoscience, 11, 727, \dodoi{10.1038/s41561-018-0233-2}

\bibitem[{Roe(2012)}]{roe_titans_2012}
Roe, H.~G. 2012, Annual Review of Earth and Planetary Sciences, 40, 355,
  \dodoi{10.1146/annurev-earth-040809-152548}

\bibitem[{Ross {et~al.}(2016)Ross, Lee, Sokol, Goldman, \&
  Bolisay}]{ross_titan_2016}
Ross, F., Lee, G., Sokol, D., Goldman, B., \& Bolisay, L. 2016, in AAS/Division
  for Planetary Sciences Meeting Abstracts \#48, Vol.~48, 123.66.
\newblock \url{http://adsabs.harvard.edu/abs/2016DPS....4812366R}

\bibitem[{Russell \& Strange(2009)}]{russell_cycler_2009}
Russell, R.~P., \& Strange, N.~J. 2009, Journal of Guidance, Control, and
  Dynamics, 32, 143, \dodoi{10.2514/1.36610}

\bibitem[{Sandstr\"{o}m \& Rahm(2020)}]{sandstrom_can_2020}
Sandstr\"{o}m, H., \& Rahm, M. 2020, Science Advances, 6, eaax0272,
  \dodoi{10.1126/sciadv.aax0272}

\bibitem[{Sciamma-O'Brien {et~al.}(2017)Sciamma-O'Brien, Upton, \&
  Salama}]{sciamma-obrien_titan_2017}
Sciamma-O'Brien, E., Upton, K.~T., \& Salama, F. 2017, Icarus, 289, 214,
  \dodoi{10.1016/j.icarus.2017.02.004}

\bibitem[{Sebree {et~al.}(2018)Sebree, Roach, Shipley, He, \&
  H\"{o}rst}]{sebree_detection_2018}
Sebree, J.~A., Roach, M.~C., Shipley, E.~R., He, C., \& H\"{o}rst, S.~M. 2018,
  The Astrophysical Journal, 865, 133, \dodoi{10.3847/1538-4357/aadba1}

\bibitem[{Sebree {et~al.}(2016)Sebree, Stern, Mandt, Domagal-Goldman, \&
  Trainer}]{sebree_13c_2016}
Sebree, J.~A., Stern, J.~C., Mandt, K.~E., Domagal-Goldman, S.~D., \& Trainer,
  M.~G. 2016, Icarus, 270, 421, \dodoi{10.1016/j.icarus.2015.04.016}

\bibitem[{Seignovert {et~al.}(2017)Seignovert, Rannou, Lavvas, Cours, \&
  West}]{seignovert_aerosols_2017}
Seignovert, B., Rannou, P., Lavvas, P., Cours, T., \& West, R.~A. 2017, Icarus,
  292, 13, \dodoi{10.1016/j.icarus.2017.03.026}

\bibitem[{Seignovert {et~al.}(2021)Seignovert, Rannou, West, \&
  Vinatier}]{seignovert_haze_2021}
Seignovert, B., Rannou, P., West, R.~A., \& Vinatier, S. 2021, The
  Astrophysical Journal, 907, 36, \dodoi{10.3847/1538-4357/abcd3b}

\bibitem[{Sharkey {et~al.}(2021)Sharkey, Teanby, Sylvestre, Mitchell, Seviour,
  Nixon, \& Irwin}]{sharkey_potential_2021}
Sharkey, J., Teanby, N.~A., Sylvestre, M., {et~al.} 2021, Icarus, 354, 114030,
  \dodoi{10.1016/j.icarus.2020.114030}

\bibitem[{Shock \& Holland(2007)}]{shock_quantitative_2007}
Shock, E.~L., \& Holland, M.~E. 2007, Astrobiology, 7, 839,
  \dodoi{10.1089/ast.2007.0137}

\bibitem[{Smith {et~al.}(2016)Smith, Cooper, \& Moores}]{smith_possible_2016}
Smith, C.~L., Cooper, B.~A., \& Moores, J.~E. 2016, Icarus, 271, 269,
  \dodoi{10.1016/j.icarus.2016.02.002}

\bibitem[{Smith {et~al.}(1996)Smith, Lemmon, Lorenz, Sromovsky, Caldwell, \&
  Allison}]{smith_titans_1996}
Smith, P.~H., Lemmon, M.~T., Lorenz, R.~D., {et~al.} 1996, Icarus, 119, 336,
  \dodoi{10.1006/icar.1996.0023}

\bibitem[{Soderblom {et~al.}(2010)Soderblom, Brown, Soderblom, Barnes, Jaumann,
  Mouélic, Sotin, Stephan, Baines, Buratti, Clark, \&
  Nicholson}]{soderblom_geology_2010}
Soderblom, J.~M., Brown, R.~H., Soderblom, L.~A., {et~al.} 2010, Icarus, 208,
  905, \dodoi{10.1016/j.icarus.2010.03.001}

\bibitem[{Soderblom {et~al.}(2007)Soderblom, Kirk, Lunine, Anderson, Baines,
  Barnes, Barrett, Brown, Buratti, Clark, Cruikshank, Elachi, Janssen, Jaumann,
  Karkoschka, Mouélic, Lopes, Lorenz, McCord, Nicholson, Radebaugh, Rizk,
  Sotin, Stofan, Sucharski, Tomasko, \& Wall}]{soderblom_correlations_2007}
Soderblom, L.~A., Kirk, R.~L., Lunine, J.~I., {et~al.} 2007, Planetary and
  Space Science, 55, 2025, \dodoi{10.1016/j.pss.2007.04.014}

\bibitem[{Soderblom {et~al.}(2009)Soderblom, Brown, Soderblom, Barnes, Kirk,
  Sotin, Jaumann, Mackinnon, Mackowski, Baines, Buratti, Clark, \&
  Nicholson}]{soderblom_geology_2009}
Soderblom, L.~A., Brown, R.~H., Soderblom, J.~M., {et~al.} 2009, Icarus, 204,
  610, \dodoi{10.1016/j.icarus.2009.07.033}

\bibitem[{Solomonidou {et~al.}(2018)Solomonidou, Coustenis, Lopes, Malaska,
  Rodriguez, Drossart, Elachi, Schmitt, Philippe, Janssen, Hirtzig, Wall,
  Sotin, Lawrence, Altobelli, Bratsolis, Radebaugh, Stephan, Brown, Mouélic,
  Gall, Villanueva, Brossier, Bloom, Witasse, Matsoukas, \&
  Schoenfeld}]{solomonidou_spectral_2018}
Solomonidou, A., Coustenis, A., Lopes, R. M.~C., {et~al.} 2018, Journal of
  Geophysical Research: Planets, 123, 489,
  \dodoi{https://doi.org/10.1002/2017JE005477}

\bibitem[{Solomonidou {et~al.}(2020{\natexlab{a}})Solomonidou, Le~Gall,
  Malaska, Birch, Lopes, Coustenis, Rodriguez, Wall, Michaelides, Nasr, Elachi,
  Hayes, Soderblom, Schoenfeld, Matsoukas, Drossart, Janssen, Lawrence,
  Witasse, Yates, \& Radebaugh}]{solomonidou_spectral_2020}
Solomonidou, A., Le~Gall, A., Malaska, M.~J., {et~al.} 2020{\natexlab{a}},
  Icarus, 344, 113338, \dodoi{10.1016/j.icarus.2019.05.040}

\bibitem[{Solomonidou {et~al.}(2020{\natexlab{b}})Solomonidou, Neish,
  Coustenis, Malaska, Le~Gall, Lopes, Werynski, Markonis, Lawrence, Altobelli,
  Witasse, Schoenfeld, Matsoukas, Baziotis, \&
  Drossart}]{solomonidou_chemical_2020}
Solomonidou, A., Neish, C., Coustenis, A., {et~al.} 2020{\natexlab{b}},
  Astronomy and Astrophysics, 641, A16, \dodoi{10.1051/0004-6361/202037866}

\bibitem[{Sotin {et~al.}(2011)Sotin, Altwegg, Brown, Hand, Lunine, Soderblom,
  Spencer, Tortora, \& {JET Team}}]{sotin_jet_2011}
Sotin, C., Altwegg, K., Brown, R.~H., {et~al.} 2011, in Lunar and Planetary
  Science Conference, Vol.~42, 1326.
\newblock \url{http://adsabs.harvard.edu/abs/2011LPI....42.1326S}

\bibitem[{Sotin {et~al.}(2012)Sotin, Lawrence, Reinhardt, Barnes, Brown, Hayes,
  Le~Mouélic, Rodriguez, Soderblom, Soderblom, Baines, Buratti, Clark,
  Jaumann, Nicholson, \& Stephan}]{sotin_observations_2012}
Sotin, C., Lawrence, K.~J., Reinhardt, B., {et~al.} 2012, Icarus, 221, 768,
  \dodoi{10.1016/j.icarus.2012.08.017}

\bibitem[{Sotin {et~al.}(2017)Sotin, Hayes, Malaska, Mastrogiuseppe, Mazarico,
  Soderblom, Tortora, Trainer, \& Turtle}]{sotin_oceanus_2017}
Sotin, C., Hayes, A., Malaska, M., {et~al.} 2017, in European Planetary Science
  Congress, Vol.~11, EPSC2017--157.
\newblock \url{http://adsabs.harvard.edu/abs/2017EPSC...11..157S}

\bibitem[{Steckloff {et~al.}(2020)Steckloff, Soderblom, Farnsworth, Chevrier,
  Hanley, Soto, Groven, Grundy, Pearce, Tegler, \&
  Engle}]{steckloff_stratification_2020}
Steckloff, J.~K., Soderblom, J.~M., Farnsworth, K.~K., {et~al.} 2020, The
  Planetary Science Journal, 1, 26, \dodoi{10.3847/PSJ/ab974e}

\bibitem[{Stevenson {et~al.}(2015)Stevenson, Lunine, \&
  Clancy}]{stevenson_membrane_2015}
Stevenson, J., Lunine, J., \& Clancy, P. 2015, Science Advances, 1, e1400067,
  \dodoi{10.1126/sciadv.1400067}

\bibitem[{Stofan {et~al.}(2013)Stofan, Lorenz, Lunine, Bierhaus, Clark,
  Mahaffy, \& Ravine}]{stofan_time_2013}
Stofan, E., Lorenz, R., Lunine, J., {et~al.} 2013, in Proceedings of the 2013
  IEEE Aerospace Conference, 211, \dodoi{10.1109/AERO.2013.6497165}

\bibitem[{Stofan {et~al.}(2007)Stofan, Elachi, Lunine, Lorenz, Stiles,
  Mitchell, Ostro, Soderblom, Wood, Zebker, Wall, Janssen, Kirk, Lopes,
  Paganelli, Radebaugh, Wye, Anderson, Allison, Boehmer, Callahan, Encrenaz,
  Flamini, Francescetti, Gim, Hamilton, Hensley, Johnson, Kelleher, Muhleman,
  Paillou, Picardi, Posa, Roth, Seu, Shaffer, Vetrella, \&
  West}]{stofan_lakes_2007}
Stofan, E.~R., Elachi, C., Lunine, J.~I., {et~al.} 2007, Nature, 445, 61,
  \dodoi{10.1038/nature05438}

\bibitem[{Stähler {et~al.}(2018)Stähler, Panning, Vance, Lorenz, van Driel,
  Nissen‐Meyer, \& Kedar}]{stahler_seismic_2018}
Stähler, S.~C., Panning, M.~P., Vance, S.~D., {et~al.} 2018, Journal of
  Geophysical Research: Planets, 123, 206,
  \dodoi{https://doi.org/10.1002/2017JE005338}

\bibitem[{Sulaiman(in review)}]{sulaiman_joint}
Sulaiman, A. in review, Experimental Astronomy

\bibitem[{Tagliabue {et~al.}(2020)Tagliabue, Schneider, Pavone, \&
  Agha-mohammadi}]{tagliabue_shapeshifter_2020}
Tagliabue, A., Schneider, S., Pavone, M., \& Agha-mohammadi, A.-a. 2020, arXiv
  e-prints, 2002, arXiv:2002.00515.
\newblock \url{http://adsabs.harvard.edu/abs/2020arXiv200200515T}

\bibitem[{Teanby {et~al.}(2019)Teanby, Sylvestre, Sharkey, Nixon, Vinatier, \&
  Irwin}]{teanby_seasonal_2019}
Teanby, N.~A., Sylvestre, M., Sharkey, J., {et~al.} 2019, Geophysical Research
  Letters, 46, 3079, \dodoi{https://doi.org/10.1029/2018GL081401}

\bibitem[{Teanby {et~al.}(2018)Teanby, Cordiner, Nixon, Irwin, H\"{o}rst,
  Sylvestre, Serigano, Thelen, Richards, \& Charnley}]{teanby_origin_2018}
Teanby, N.~A., Cordiner, M.~A., Nixon, C.~A., {et~al.} 2018, The Astronomical
  Journal, 155, 251, \dodoi{10.3847/1538-3881/aac172}

\bibitem[{Telfer {et~al.}(2019)Telfer, Radebaugh, Cornford, \&
  Lewis}]{telfer_long-wavelength_2019}
Telfer, M.~W., Radebaugh, J., Cornford, B., \& Lewis, C. 2019, Journal of
  Geophysical Research: Planets, 124, 2369,
  \dodoi{https://doi.org/10.1029/2019JE006117}

\bibitem[{Thelen {et~al.}(2019{\natexlab{a}})Thelen, Nixon, Cordiner, Charnley,
  Irwin, \& Kisiel}]{thelen_measurement_2019}
Thelen, A.~E., Nixon, C.~A., Cordiner, M.~A., {et~al.} 2019{\natexlab{a}}, The
  Astronomical Journal, 157, 219, \dodoi{10.3847/1538-3881/ab19bb}

\bibitem[{Thelen {et~al.}(2019{\natexlab{b}})Thelen, Nixon, Chanover, Cordiner,
  Molter, Teanby, Irwin, Serigano, \& Charnley}]{thelen_abundance_2019}
Thelen, A.~E., Nixon, C.~A., Chanover, N.~J., {et~al.} 2019{\natexlab{b}},
  Icarus, 319, 417, \dodoi{10.1016/j.icarus.2018.09.023}

\bibitem[{Thelen {et~al.}(2020)Thelen, Cordiner, Nixon, Vuitton, Kisiel,
  Charnley, Palmer, Teanby, \& Irwin}]{thelen_detection_2020}
Thelen, A.~E., Cordiner, M.~A., Nixon, C.~A., {et~al.} 2020, The Astrophysical
  Journal, 903, L22, \dodoi{10.3847/2041-8213/abc1e1}

\bibitem[{Tobie {et~al.}(2012)Tobie, Gautier, \&
  Hersant}]{tobie_titantextquotesingles_2012}
Tobie, G., Gautier, D., \& Hersant, F. 2012, The Astrophysical Journal, 752,
  125, \dodoi{10.1088/0004-637X/752/2/125}

\bibitem[{Tobie {et~al.}(2006)Tobie, Lunine, \& Sotin}]{tobie_episodic_2006}
Tobie, G., Lunine, J.~I., \& Sotin, C. 2006, Nature, 440, 61,
  \dodoi{10.1038/nature04497}

\bibitem[{Tokano(2009)}]{tokano_limnological_2009}
Tokano, T. 2009, Astrobiology, 9, 147, \dodoi{10.1089/ast.2007.0220}

\bibitem[{Tokano(2015)}]{tokano_precipitation_2015}
---. 2015, Origins of Life and Evolution of Biospheres, 45, 231,
  \dodoi{10.1007/s11084-015-9424-7}

\bibitem[{Tokano(2020)}]{tokano_stable_2020}
---. 2020, Geophysical Research Letters, 47, e2019GL086166,
  \dodoi{https://doi.org/10.1029/2019GL086166}

\bibitem[{Tokano \& Lorenz(2019)}]{tokano_modeling_2019}
Tokano, T., \& Lorenz, R.~D. 2019, Journal of Geophysical Research: Planets,
  124, 617, \dodoi{https://doi.org/10.1029/2018JE005898}

\bibitem[{Toledo {et~al.}(2019)Toledo, Irwin, Rannou, Teanby, Simon, Wong, \&
  Orton}]{toledo_constraints_2019}
Toledo, D., Irwin, P. G.~J., Rannou, P., {et~al.} 2019, Icarus, 333, 1,
  \dodoi{10.1016/j.icarus.2019.05.018}

\bibitem[{Tortora {et~al.}(2018)Tortora, Buccino, Oudrhiri, Zannoni,
  Gomez~Casajus, Mitri, \& Lombardo}]{tortora_ocean_2018}
Tortora, P., Buccino, D., Oudrhiri, K., {et~al.} 2018, in 42nd COSPAR
  Scientific Assembly, Vol.~42, B0.2--12--18.
\newblock \url{http://adsabs.harvard.edu/abs/2018cosp...42E3416T}

\bibitem[{Trainer(2013)}]{g_trainer_atmospheric_2013}
Trainer, M.~G. 2013, Current organic chemistry, 17, 1710

\bibitem[{Trainer {et~al.}(2006)Trainer, Pavlov, DeWitt, Jimenez, McKay, Toon,
  \& Tolbert}]{trainer_organic_2006}
Trainer, M.~G., Pavlov, A.~A., DeWitt, H.~L., {et~al.} 2006, Proceedings of the
  National Academy of Sciences, 103, 18035, \dodoi{10.1073/pnas.0608561103}

\bibitem[{Turtle {et~al.}(2011{\natexlab{a}})Turtle, Genio, Barbara, Perry,
  Schaller, McEwen, West, \& Ray}]{turtle_seasonal_2011}
Turtle, E.~P., Genio, A. D.~D., Barbara, J.~M., {et~al.} 2011{\natexlab{a}},
  Geophysical Research Letters, 38,
  \dodoi{https://doi.org/10.1029/2010GL046266}

\bibitem[{Turtle {et~al.}(2009)Turtle, Perry, McEwen, DelGenio, Barbara, West,
  Dawson, \& Porco}]{turtle_cassini_2009}
Turtle, E.~P., Perry, J.~E., McEwen, A.~S., {et~al.} 2009, Geophysical Research
  Letters, 36, \dodoi{https://doi.org/10.1029/2008GL036186}

\bibitem[{Turtle {et~al.}(2011{\natexlab{b}})Turtle, Perry, Hayes, Lorenz,
  Barnes, McEwen, West, Genio, Barbara, Lunine, Schaller, Ray, Lopes, \&
  Stofan}]{turtle_rapid_2011}
Turtle, E.~P., Perry, J.~E., Hayes, A.~G., {et~al.} 2011{\natexlab{b}},
  Science, 331, 1414, \dodoi{10.1126/science.1201063}

\bibitem[{Turtle {et~al.}(2017)Turtle, Barnes, Trainer, Lorenz, MacKenzie,
  Hibbard, Adams, Bedini, Langelaan, Zacny, \& {Dragonfly
  Team}}]{turtle_dragonfly_2017}
Turtle, E.~P., Barnes, J.~W., Trainer, M.~G., {et~al.} 2017, in Lunar and
  {Planetary} {Science} {Conference}, Lunar and {Planetary} {Science}
  {Conference}, 1958

\bibitem[{Turtle {et~al.}(2018)Turtle, Perry, Barbara, Genio, Rodriguez,
  Mouélic, Sotin, Lora, Faulk, Corlies, Kelland, MacKenzie, West, McEwen,
  Lunine, Pitesky, Ray, \& Roy}]{turtle_titans_2018}
Turtle, E.~P., Perry, J.~E., Barbara, J.~M., {et~al.} 2018, Geophysical
  Research Letters, 45, 5320, \dodoi{https://doi.org/10.1029/2018GL078170}

\bibitem[{Vance {et~al.}(2018)Vance, Panning, Stähler, Cammarano, Bills,
  Tobie, Kamata, Kedar, Sotin, Pike, Lorenz, Huang, Jackson, \&
  Banerdt}]{vance_geophysical_2018}
Vance, S.~D., Panning, M.~P., Stähler, S., {et~al.} 2018, Journal of
  Geophysical Research: Planets, 123, 180,
  \dodoi{https://doi.org/10.1002/2017JE005341}

\bibitem[{Vinatier {et~al.}(2012)Vinatier, Rannou, Anderson, Bézard, de~Kok,
  \& Samuelson}]{vinatier_optical_2012}
Vinatier, S., Rannou, P., Anderson, C.~M., {et~al.} 2012, Icarus, 219, 5,
  \dodoi{10.1016/j.icarus.2012.02.009}

\bibitem[{Vinatier {et~al.}(2018)Vinatier, Schmitt, Bézard, Rannou, Dauphin,
  de~Kok, Jennings, \& Flasar}]{vinatier_study_2018}
Vinatier, S., Schmitt, B., Bézard, B., {et~al.} 2018, Icarus, 310, 89,
  \dodoi{10.1016/j.icarus.2017.12.040}

\bibitem[{Vinatier {et~al.}(2020)Vinatier, Mathé, Bézard, d’Ollone,
  Lebonnois, Dauphin, Flasar, Achterberg, Seignovert, Sylvestre, Teanby,
  Gorius, Mamoutkine, Guandique, \& Jennings}]{vinatier_temperature_2020}
Vinatier, S., Mathé, C., Bézard, B., {et~al.} 2020, Astronomy \&
  Astrophysics, 641, A116, \dodoi{10.1051/0004-6361/202038411}

\bibitem[{Vu {et~al.}(2020)Vu, Choukroun, Sotin, Muñoz‐Iglesias, \&
  Maynard‐Casely}]{vu_rapid_2020}
Vu, T.~H., Choukroun, M., Sotin, C., Muñoz‐Iglesias, V., \&
  Maynard‐Casely, H.~E. 2020, Geophysical Research Letters, 47,
  e2019GL086265, \dodoi{https://doi.org/10.1029/2019GL086265}

\bibitem[{Vuitton {et~al.}(2019)Vuitton, Yelle, Klippenstein, H\"{o}rst, \&
  Lavvas}]{vuitton_simulating_2019}
Vuitton, V., Yelle, R.~V., Klippenstein, S.~J., H\"{o}rst, S.~M., \& Lavvas, P.
  2019, Icarus, 324, 120, \dodoi{10.1016/j.icarus.2018.06.013}

\bibitem[{Vuitton {et~al.}(2007)Vuitton, Yelle, \& McEwan}]{vuitton_ion_2007}
Vuitton, V., Yelle, R.~V., \& McEwan, M.~J. 2007, Icarus, 191, 722,
  \dodoi{10.1016/j.icarus.2007.06.023}

\bibitem[{Waite {et~al.}(2013)Waite, Bell, Lorenz, Achterberg, \&
  Flasar}]{waite_model_2013}
Waite, J.~H., Bell, J., Lorenz, R., Achterberg, R., \& Flasar, F.~M. 2013,
  Planetary and Space Science, 86, 45, \dodoi{10.1016/j.pss.2013.05.018}

\bibitem[{Waite {et~al.}(2007)Waite, Young, Cravens, Coates, Crary, Magee, \&
  Westlake}]{waite_process_2007}
Waite, J.~H., Young, D.~T., Cravens, T.~E., {et~al.} 2007, Science, 316, 870,
  \dodoi{10.1126/science.1139727}

\bibitem[{Wasiak {et~al.}(2013)Wasiak, Androes, Blackburn, Tullis, Dixon, \&
  Chevrier}]{wasiak_geological_2013}
Wasiak, F.~C., Androes, D., Blackburn, D.~G., {et~al.} 2013, Planetary and
  Space Science, 84, 141, \dodoi{10.1016/j.pss.2013.05.007}

\bibitem[{Wellbrock {et~al.}(2013)Wellbrock, Coates, Jones, Lewis, \&
  Waite}]{wellbrock_cassini_2013}
Wellbrock, A., Coates, A.~J., Jones, G.~H., Lewis, G.~R., \& Waite, J.~H. 2013,
  Geophysical Research Letters, 40, 4481,
  \dodoi{https://doi.org/10.1002/grl.50751}

\bibitem[{Wellbrock {et~al.}(2019)Wellbrock, Coates, Jones, Vuitton, Lavvas,
  Desai, \& Waite}]{wellbrock_heavy_2019}
Wellbrock, A., Coates, A.~J., Jones, G.~H., {et~al.} 2019, Monthly Notices of
  the Royal Astronomical Society, 490, 2254, \dodoi{10.1093/mnras/stz2655}

\bibitem[{Werynski {et~al.}(2019)Werynski, Neish, Gall, \&
  Janssen}]{werynski_compositional_2019}
Werynski, A., Neish, C.~D., Gall, A.~L., \& Janssen, M.~A. 2019, Icarus, 321,
  508, \dodoi{10.1016/j.icarus.2018.12.007}

\bibitem[{West {et~al.}(2016)West, Del~Genio, Barbara, Toledo, Lavvas, Rannou,
  Turtle, \& Perry}]{west_cassini_2016}
West, R.~A., Del~Genio, A.~D., Barbara, J.~M., {et~al.} 2016, Icarus, 270, 399,
  \dodoi{10.1016/j.icarus.2014.11.038}

\bibitem[{West {et~al.}(2018)West, Seignovert, Rannou, Dumont, Turtle, Perry,
  Roy, \& Ovanessian}]{west_seasonal_2018}
West, R.~A., Seignovert, B., Rannou, P., {et~al.} 2018, Nature Astronomy, 2,
  495, \dodoi{10.1038/s41550-018-0434-z}

\bibitem[{Wong {et~al.}(2015)Wong, Yung, \&
  Randall~Gladstone}]{wong_plutos_2015}
Wong, M.~L., Yung, Y.~L., \& Randall~Gladstone, G. 2015, Icarus, 246, 192,
  \dodoi{10.1016/j.icarus.2014.05.019}

\bibitem[{Woodson {et~al.}(2015)Woodson, Smith, Crary, \&
  Johnson}]{woodson_ion_2015}
Woodson, A.~K., Smith, H.~T., Crary, F.~J., \& Johnson, R.~E. 2015, Journal of
  Geophysical Research: Space Physics, 120, 212,
  \dodoi{https://doi.org/10.1002/2014JA020499}

\bibitem[{Yu {et~al.}(2020)Yu, H\"{o}rst, He, \& McGuiggan}]{yu_single_2020}
Yu, X., H\"{o}rst, S.~M., He, C., \& McGuiggan, P. 2020, Earth and Planetary
  Science Letters, 530, 115996, \dodoi{10.1016/j.epsl.2019.115996}

\bibitem[{Yu {et~al.}(2018)Yu, H\"{o}rst, He, McGuiggan, \&
  Crawford}]{yu_where_2018}
Yu, X., H\"{o}rst, S.~M., He, C., McGuiggan, P., \& Crawford, B. 2018, Journal
  of Geophysical Research: Planets, 123, 2310,
  \dodoi{https://doi.org/10.1029/2018JE005651}

\end{thebibliography}
\bibliographystyle{aasjournal}


\end{document}